\documentclass[useAMS,usenatbib,usegraphicx,referee, amsthm]{biom}
\usepackage{amssymb}
\usepackage{tikz}
\usetikzlibrary{arrows, calc, decorations.pathmorphing, backgrounds, positioning, fit, petri, automata}
\usepackage{caption}
\usepackage{graphicx}
\usepackage{subfigure}
\usepackage{rotating}
\usepackage{chngcntr}
\usepackage{apptools}
\usepackage{comment}

\usepackage{hyperref}
\usepackage{xcolor}         %for colors
\definecolor{grau}{rgb}{0.8,0.8,0.8}

\newcommand{\chen}[1]{\color{orange}}
\usepackage{setspace,lscape,longtable}
\usepackage{graphicx}
\usepackage{amsmath}
\usepackage{algorithm}
\usepackage{algpseudocode}
\usepackage[para,online,flushleft]{threeparttable}
\usepackage{siunitx}

\newcommand{\bY}{\bm{Y}}

\newcommand{\bZ}{\bm{Z}}
\newcommand{\xb}{\bm{x}}
\newcommand{\bmu}{\bm{\mu}}

\newcommand{\bc}{\bm{c}}

\newcommand{\balpha}{\bm{\alpha}}

\newcommand{\bgamma}{\bm{\gamma}}

\newcommand{\iid}{\stackrel{iid}{\sim}}

\title[Bayesian Enrichment Design]{ASIED: A Bayesian Adaptive Subgroup-Identification Enrichment Design}

\author{Yanxun Xu$^{1,*}$\email{yanxun.xu@jhu.edu, 410-516-7341}, Florica Constantine$^{1}$, Yuan Yuan$^{2}$, Yili L. Pritchett$^{2,3}$ \\
$^{1}$Department of Applied Mathematics and Statistics, Johns Hopkins University, U.S.A. \\
$^{2}$MedImmune, LLC, U.S.A. \\
$^{3}$Present affiliate: G1 Therapeutics, Inc., U.S.A.
}

\begin{document}
%  This will produce the submission and review information that appears
%  right after the reference section.  Of course, it will be unknown when
%  you submit your paper, so you can either leave this out or put in
%  sample dates (these will have no effect on the fate of your paper in the
%  review process!)

%\date{{\it Received April} 2007. {\it Revised April} 2007.  {\it
%Accepted April} 2007.}

%\date{{\it } 2017. {\it } 2007.  {\it } 2017.}

\pagerange{\pageref{firstpage}--\pageref{lastpage}}
\volume{}
\pubyear{}
\artmonth{}

\doi{}

\label{firstpage}

\begin{abstract}
Developing targeted therapies based on patients' baseline characteristics and genomic profiles such as biomarkers has gained growing interests in recent years. 
Depending on  patients' clinical characteristics, the expression of specific biomarkers or their combinations, different patient subgroups could respond differently to the same treatment. An ideal design, especially at the proof of concept stage, should search for such subgroups and make dynamic adaptation as the trial goes on. 
When no prior knowledge is available on whether the treatment works on the all-comer population or only works on the subgroup defined by one biomarker or several biomarkers, it's necessary to incorporate the adaptive estimation of the heterogeneous treatment effect to the decision-making 
%estimate the subgroup effect adaptively based on the response outcomes and biomarker profiles from all the treated subjects
at interim analyses. 
To address this problem, we propose an Adaptive Subgroup-Identification Enrichment Design, ASIED,  to simultaneously search for predictive biomarkers, identify the subgroups with differential treatment effects, and modify study entry criteria at  interim analyses when justified.  More importantly, we construct robust quantitative decision-making rules for population enrichment when the interim outcomes are heterogeneous in the context of a multilevel target product profile, which defines the minimal and targeted levels of treatment effect. 
 Through extensive simulations, the ASIED is demonstrated to achieve desirable operating characteristics and compare favorably against alternatives.
\end{abstract}

\begin{keywords}
Adaptive enrichment design, Bayesian subgroup identification, Biomarker, Decision-making, Multilevel target product profile.\\
\vskip .3in
\end{keywords}

\maketitle

\newpage
\section{Introduction}

Decision-making is a critical step for early-phase drug development, aiming to advance promising drugs for further development and stop inferior drugs. Traditionally, decisions are usually based on a significant $p$-value, e.g., the rejection of the null hypothesis in a Simon two-stage design \citep{simon1989optimal} would warrant further development of the drug. However, a rejection of the null hypothesis may not necessarily support a desired treatment effect  \citep{ratain2007testing}. Furthermore, in the context of a multilevel target product profile (TPP), which defines the minimal and targeted levels of treatment effect, more complex decisions are needed.  Such a TPP framework has been used routinely throughout clinical development since it addresses the challenges when a drug effect cannot be represented well by a single threshold  
\citep{frewer2016decision,pulkstenis2017bayesian}.  
%have shown that such a TPP framework that has been used routinely throughout clinical development can address the challenges when a drug effect cannot be represented well by a single threshold.  
\cite{lalonde2007model} first formulated a Go/Pause/Stop decision framework by defining a lower reference value (LRV) and a target value (TV), where LRV represents a ``dignity" line for developing a drug and TV represents a desired clinical improvement of commercial viability. \cite{frewer2016decision} illustrated how the Go/Pause/Stop decisions  in the paper of \cite{lalonde2007model} were calculated for a Phase II study and how the operating characteristics were assessed to ensure   robustness.  
\cite{pulkstenis2017bayesian} extended the Lalonde framework by allowing to incorporate historical data prior to calculating probabilities of Go/Pause/Stop.  However, these approaches were developed based on the assumption that the treatment effect was homogeneous among patients, limiting their applicability in many diseases where heterogeneous treatment effect might exist. 
%Applying the Lalonde framework to construct enrichment decision rules is challenging and complicated, since one needs to deal with heterogeneity in treatment effects across patient subpopulations. To our best knowledge, there is no such precedent thus far.

Depending on patients' clinical characteristics, the expression of specific biomarkers or their combinations, it has become well known that there exists heterogeneity in treatment effects across 
patient subpopulations when given the same treatment in many diseases. 
Thus, it is essential to take into account potentially heterogeneous treatment effects  in clinical trial designs or data analyses when making a decision.   
For instance, breast cancer patients with an enriched HER2 pathway were found to respond well to the medication trastuzumab \citep{hudis2007trastuzumab}, 
%through pairing genetic traits with targeted treatment options
while other subtypes of breast cancers do not. Another example is that treatments with EGFR antibodies are not recommended for KRAS mutated colorectal cancer patients as 
they are usually resistant to anti-EGFR treatment \citep{misale2012emergence}. Therefore, it is very important at the proof-of-concept stage of drug development 
to identify the biomarkers that have interaction effects with the treatment
 and are predictive of the subgroups that are more likely to respond to the treatment.

For situations where the predictive biomarkers are hypothesized but not proved, i.e.,  it is not clear whether the new therapy works for all-comers or a subpopulation, 
adaptive enrichment designs have been developed to evaluate the treatment effect and thereby 
modify the eligibility criteria at interim analyses in an attempt to identify the right population for the new therapy. 
Patients exhibiting the desired treatment effects are referred to as the ``enriched population." \cite{wang2007approaches} and  \cite{karuri2012two} 
 compared the active versus placebo by evaluating the treatment effectiveness of biomarker positive- and negative- subgroups at an interim analysis, allowing for terminating the enrollment of the biomarker negative subgroup. % \citep{wang2007approaches,karuri2012two}.  
%\cite{rosenblum2011optimizing} proposed an adaptive design in which patient enrollment may be changed based on certain pre-planned criteria assuming no data-dependent period effects. 
Simon et al. developed a class of adaptive enrichment designs to adaptively update the eligibility criteria,  they then   extended this work by developing a frequentist/Bayesian model for decision making using a formal hypothesis test at the end of the trial to preserve type I error \citep{simon2013adaptive,simon2017using}. 
See, for example, Wang and Huang for a review of adaptive enrichment designs \citep{wang2013adaptive}. 

Most of these enrichment designs use a set of biomarkers to pre-define subgroups and then test if there are differential therapeutic effects on these pre-defined subgroups. However, pre-defining subgroups can be problematic if the pre-defined biomarkers are not predictive or the cutoff values for the predictive biomarkers are incorrect. 
 Therefore, an enrichment design that allows the discovery and estimation of subgroups during the clinical trial is highly desirable. Subgroup identification involves two major tasks: identify predictive biomarkers and determine the optimal cutoff values for predictive biomarkers. 
%Sivaganesan et al. identified subgroups through a model selection procedure  to determine the presence (or absence) of treatment-subgroup interactions  \citep{sivaganesan2011bayesian}. 
 \cite{foster2011subgroup} developed a random forest-based algorithm to find subgroups by searching biomarker regions where the treatment effect is larger than the average effect on the whole population. 
\cite{lipkovich2011subgroup} developed SIDES (subgroup identification based on differential effect search), which makes use of the classification and regression tree (CART) \citep{breiman1984classification} to recursively split a patient set such that one of the halves from each split has  maximal treatment effect relative to the other half. 
\cite{loh2015regression} and \cite{loh2016identification}  proposed a regression tree approach, GUIDE, to first decide which biomarkers to split on through the use of $\chi^2$ tests, and then identify the optimal cutoff values for the selected biomarkers. \cite{xu2014subgroup} developed SUBA, a Bayesian subgroup-based adaptive design, to allocate the patients to their superior treatments using a random partition model that splits the biomarker space by the observed biomarker's median value, 
which generally is not the optimal cutoff for a predictive biomarker.  \cite{guo2017subgroup} extended SUBA ({\it i.e.}, SCUBA) by allowing the biomarker space to be split using hyperplanes that construct linear boundaries, providing a more flexible partition model. 
%Shen and He used a structured logistic-normal mixture model to test for the existence of subgroups by a confirmatory statistical test  \citep{shen2015inference}.  
 All of these methods either target the subgroup identification using the retrospective clinical trial data, or focus on subgroup identification for patient allocations during the trial. They have not been directly utilized in 
 clinical trial enrichment designs to modify the study entry criteria during interim analyses in the context of a multilevel TPP.

Applying the Lalonde framework \citep{lalonde2007model} in the context of a multilevel TPP to construct enrichment decision rules in a heterogenous population is challenging and complicated, since one needs to deal with heterogeneity in treatment effect across patient subpopulations. To our best knowledge, there is no such precedent thus far.
In this paper, we propose an adaptive subgroup-identification enrichment design (ASIED), utilizing patients' biomarker profiles and outcomes as they become available to create a design with learning, adapting and enriching capacity.  
ASIED searches for subgroups among a set of biomarkers using interim accumulated data obtained from all-comers rather than predefining subgroups and allows the entry criteria to be modified to enroll  subjects with specific clinical characteristic or biomarker signature 
who are more likely to respond to the treatment, enabling better learning of the treatment effect on the enriched population. 
 The biomarker-defined subgroups are not fixed upfront -- we assume a prior on the partition to classify the patients into  subgroups and then learn the cutoff values using observed response data. 
Building upon \cite{xu2014subgroup} and \cite{guo2017subgroup}, ASIED uses a flexible Bayesian model as 
 the core search algorithm that can handle biomarkers of varying forms (continuous, binary, categorical, or ordinal) and different types of  response  outcomes (binary, categorical, continuous). The method can be easily extended to model other types of outcomes, such as counting and survival outcomes. More importantly, we construct 
  a decision-making framework to make informed decisions, such as continuing with all-comers or enriching to a subpopulation. 
  The possible decisions include: a ``Go" decision for all-comers, a ``Go" decision for the identified subgroup with enhanced treatment effect, a ``Stop" decision for all-comers due to futility, or conduct one more interim analysis when the interim result is  inconclusive.

The key novelty of ASIED is to combine the tasks  of ``subgroup identification"  and  ``enrichment" to  create an adaptive enrichment design engine that enables more efficient learning about the investigational drug's efficacy  in the identified subpopulation through a set of  decision-making rules in the context of a multilevel TPP.  The posterior inference for subgroups with enhanced treatment effects under the proposed Bayesian model will lead to the modification of study enrollment criteria. Therefore,  more patients with the characteristics in the identified subgroup can be enrolled to the study as the trial continues after the interim analysis. 
%the subgroups with enhanced treatment effects are continuously identified and redefined based on patients' responses  at each interim analysis. The posterior inference for subgroups with enhanced treatment effects will lead to the modification of study enrollment criteria. Therefore,  more patients with the characteristics in the identified subgroup can be enrolled to the study as the trial continues after the interim analysis. This design combines the tasks  of ``subgroup identification"  and  ``enrichment" to  create an adaptive enrichment design engine that enables more efficient learning about 
% the investigational drug's efficacy  in the identified subpopulation through a set of  decision-making rules. 
In summary, ASIED represents the first effort in the literature to conduct subgroup discovery, evaluation, and adaptive modification of study enrollment criteria simultaneously using  a flexible Bayesian searching algorithm and a probabilistic decision-making framework. 

This paper proceeds as follows. We first introduce the motivating trial in Section \ref{sec:motivate}. The proposed ASIED  design with a decision-making framework is described in Section \ref{sec:design}. 
We elaborate the proposed Bayesian random partition model for subgroup identification in Section \ref{sec:model} with simulation studies.  Section \ref{sec:oc} presents  simulation studies that examine the operating characteristic of the ASIED in  the motivating AD symptom improvement POC study. We conclude with a discussion in Section \ref{sec:con}.

\section{Motivating Trial}
\label{sec:motivate}
We consider a placebo-controlled, double-blind proof-of-concept (POC) study for an investigational drug on patients with Alzheimer's disease (AD)  for symptom improvement. The primary efficacy endpoint is the change from baseline to final observation on the total score of 13-item Alzheimer's disease Assessment Scale -- Cognitive Subscale (ADAS-cog) 
%\rr (total score) 
in a 12-week study. Previous research has suggested a number of biomarkers that might predict treatment effect on AD. These biomarkers are apolipoprotein E (APOE)-$\epsilon$4 genotype and allele status, plasma amyloid precursor protein $\beta$ (A$\beta$), and cerebrospinal fluid (CSF) $\beta$-site amyloid precursor protein (APP)-cleaving enzyme 1 (BACE1). It is plausible that the investigational drug only has a clinically meaningful effect on a subpopulation that is qualified by one of the biomarkers or a combination of several biomarkers listed above. The objective of this clinical trial is to test whether the investigational drug has efficacy of AD symptom improvement on all-comers or on a biomarker-defined subpopulation, and if latter, to learn more about the efficacy  of the drug on the subpopulation in the same study by adaptive enrichment. % so that development speed for the molecular can be assumed.(to delete: combine the task of subgroup identification and population enrichment with) the purpose of developing this investigational drug more rapidly.)\xx  

At the beginning, patients diagnosed with probable AD and 
meeting entry criteria will be enrolled and equally assigned to the placebo or the investigational drug. Baseline biomarkers data will be collected. At the pre-specified interim analysis, accumulating ADAS-cog total scores are utilized to assess whether the treatment has effect on all-comers and  search for potential subgroups with enhanced treatment effects compared to all-comers. When such a subgroup is ascertained by pre-specified decision rules, study entry criteria will be modified so that only the patients with the characteristics in the identified subpopulation will be enrolled  in the rest of the study. 
% Sample size estimation for detecting the treatment effect on the subpopulation or adding more doses to the enriched part of the design are options for consideration as part of the enrichment plan. \xx
The enrichment in the middle course of the study  will allow  
more information  to be obtained for the subpopulation to inform next step of clinical development. %\rr should the treatment only has efficacy on a sub-population but not on all-comers. \xx 

%Based on the results from 10 placebo-controlled studies on Cholinesterase Inhibitors (approved AD symptom improvement therapies) reported by \citep{birks2006cholinesterase}, 
 Previous research reported that the pooled treatment effect of Cholinesterase Inhibitors (approved AD symptom improvement therapies) vs. placebo on mild-to-moderate AD patients was 2.37 \citep{birks2006cholinesterase}. 
%with standard deviation around 6.0. 
In this paper, the motivating AD trial is used as the background for simulation studies. To set up simulation scenarios and Go/No-Go decision rules in the proposed adaptive enrichment design, we assume that LRV=2.37 is the minimum target and a 30\% gain over the pooled effect (TV=3.08) is the desired target when developing a new molecular for the AD symptom treatment. %In the motivating trial design, we will use total sample size of $N=180$ and conduct first interim when $n_1=100$ (50/arm) have finished Week 12 visit. 

\section{Adaptive Subgroup-Identification Enrichment Design (ASIED) }
\label{sec:design}
Assume we have a maximum sample size of $N$ patients and $T$ candidate treatments  indexed by $t = 1, \ldots, T$. In the motivating AD trial, $T=2$, where $t=1$ represents the placebo and $t=2$ represents the investigational drug. All patients are equally randomized to placebo and investigational drug. 
The design can be easily extended to multiple treatments. 
In this section, we propose an adaptive enrichment design that applies a set of decision rules to determine whether the study population should be enriched after subgroup identification. 
%uses BayRP as the core subgroup searching engine and 
%Without loss of generality we assume that  the outcome is continuous and  $T=2$ as in the motivating AD trial, where $t=1$ represents the placebo and $t=2$ represents the investigational drug. 
%% All patients are equally randomized to placebo and investigational drug. 
%The design can be easily extended to multiple treatments. 

In the motivating AD trial, we have 
a lower reference value (LRV) and a target value (TV) to characterize the desired efficacy, in which LRV represents a clinically meaningful minimum treatment effect that is a ``dignity" line for developing a drug and  TV represents a desired targeted effect increment of the new drug ($\mathrm{TV}>\mathrm{LRV})$.    Let $\delta_{\Delta}$ represent the estimated treatment effect for a given subgroup $\Delta$ under a probability model. 
Note here we do not make any specific assumptions about the subgroup identification model, except for the existence of such a model.  
Then the probability of the estimated treatment effect  
being larger than LRV for a given subgroup $\Delta$ can be denoted by $Pr(\delta_{\Delta}\geq \mathrm{LRV})$. 
%by 
%$$Pr(\delta_{\Delta}\geq \mathrm{LRV}) = \frac{1}{B}\sum_bI\big[\delta^{(b)}_{\Delta}>\mathrm{LRV}\big].$$ 
Similarly for the probability of the estimated treatment effect  
being larger than TV for a given subgroup: $Pr(\delta_{\Delta}\geq \mathrm{TV})$. 

We define three decision outcomes for a subgroup $\Delta$: ``Go" if $Pr(\delta_{\Delta}\geq \mathrm{LRV})\geq \xi_1$; ``Stop" if $Pr(\delta_{\Delta}\geq \mathrm{LRV})< \xi_1$ and $Pr(\delta_{\Delta}\geq \mathrm{TV})< \xi_2$; ``Gray zone" otherwise. The tuning parameters  $\xi_1$ and $\xi_2$  are chosen by simulations to obtain a design with desirable  operating characteristics. In Section \ref{sec:oc}, we will illustrate how one may calibrate these parameters.

The ASIED will be conducted as follows. %Assume that the maximum sample size is $N$. 
\begin{itemize}
\item Start the trial by enrolling all-comers denoted by $\Omega$.  All  subjects are equally randomized to the placebo and the investigational drug. 
\item \underline{Interim analyses.} At the time when the data from the first $n_1$ subjects become available, an interim analysis will be conducted.  Specifically, we 
%implement the proposed BayRP to compute the posterior distribution of parameters of interests, 
search for subgroups with enhanced treatment effect, and apply the following rules to make decisions.
\begin{enumerate}
\item If the treatment effect is a ``Go" for all-comers, that is, $Pr(\delta_{\Omega}\geq \mathrm{LRV})\geq \xi_1$, we continue the trial with the all-comer population until $N$ subjects are enrolled. 
\item If the treatment effect is a ``Stop" for all-comers, but a ``Go" for a subgroup $\Delta$, that is, $Pr(\delta_{\Omega}\geq \mathrm{LRV})< \xi_1$, $Pr(\delta_{\Omega}\geq \mathrm{TV})< \xi_2$, and $Pr(\delta_{\Delta}\geq \mathrm{LRV})\geq \xi_1$, 
we enrich the study population to this subgroup by restricting the entry into the clinical trial to only patients with $\xb\in \Delta$ for the remaining sample size (e.g., $N-n_1$).  Here $\xb\in \Delta$ indicates that the patient with biomarker profile $\xb$ belongs to the subgroup $\Delta$. 
\item If the treatment effect is a ``Stop" for all-comers and all possible subgroups, that is, $Pr(\delta_{\Omega}\geq \mathrm{LRV})< \xi_1$, $Pr(\delta_{\Omega}\geq \mathrm{TV})< \xi_2$, $Pr(\delta_{\Delta}\geq \mathrm{LRV})< \xi_1$, and $Pr(\delta_{\Delta}\geq \mathrm{TV})< \xi_2$ for any subgroup $\Delta$, then we stop the trial early due to futility. 
\item If the treatment effect for all-comers is in ``Gray zone", that is, $Pr(\delta_{\Omega}\geq \mathrm{LRV})< \xi_1$ and  $Pr(\delta_{\Omega}\geq \mathrm{TV})\geq \xi_2$, we continue the trial with the all-comer population with $n_2$ ($n_1+n_2<N$) more patients and conduct a second interim analysis with potential outcomes of Step 1, 2, or 3 at the second interim. 
\item If the treatment effect for all-comers is ``Stop", but there is one subgroup $\Delta$ in ``Gray zone", we enrich study population to this subgroup by restricting the entry into the clinical trial to only patients with $\xb\in \Delta$ until $n_2$ more patients are enrolled. Then we conduct a second interim analysis with the following potential outcomes. 
\begin{itemize}
\item If the treatment effect is a ``Stop" for the enriched subpopulation $\Delta$, we stop the trial due to futility. 
\item Otherwise, we continue the trial with the enriched subpopulation $\Delta$ until all N subjects are enrolled to the study. 
\end{itemize}
\end{enumerate}

\item \underline{Final   recommendation of the drug.} When the trial is completed, there are three possible final  recommendations of the drug denoted by  $a\in\{0, 1, 2\}$, with $a=2$ denoting a recommendation of ``Go" for all-comers; $a=1$ denoting a recommendation of ``Go" for a subgroup; otherwise $a=0$  (including the scenarios where the investigational drug fails to demonstrate effect for all-comers nor for any sub-population, the treatment effect is in the "Gray Zone" for all-comers, or it is ``Stop"  for all-comers but in the "Gray Zone" for a subgroup). 

\end{itemize}

\begin{figure}[h]
\centering
\includegraphics[scale=0.45]{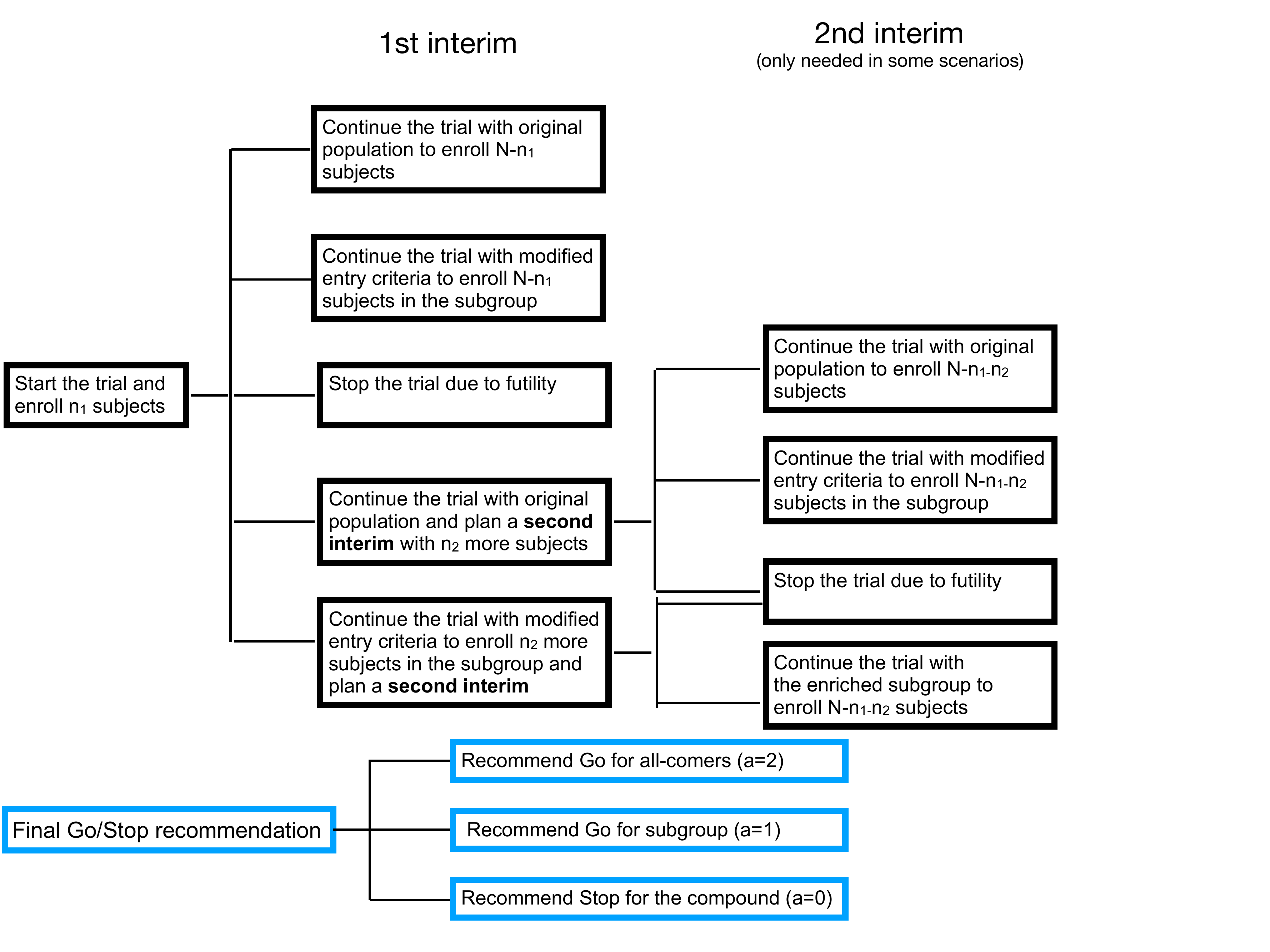}
\caption{A flowchart of the ASIED decision-making framework. The boxes in blue color represent the final recommendation of Go/Stop. }
\label{fig:trial}
\end{figure}

%and $a=0$ denoting recommendation of ``Stop". Here ``Stop" indicates that the investigational drug is ineffective and will not be allowed  to go on to a later phase. 
%\rr (Suggest to delete the following) If the trial is stopped early due to futility, we record the terminal decision $a=0$. When the trial reaches the maximum number of patients, if the treatment effect is a ``Go" for all-comers, we record the final decision $a=2$; if the treatment effect is a ``Stop" for all-comers but a ``Go" for a subgroup $\Delta$, we record the final decision $a=1$; otherwise $a=0$. \xx
Figure \ref{fig:trial} summarizes the decision-making framework for the ASIED. 
%\rr (Suggest to add number to each decision outcome ($1,2, ..,5$). With these, the 2nd interim outcomes can be denoted as $1,2,3$, and a new one, $6$. Also,since as we tried to clarify that a=0 meanings several things, including stopping the drug's further development and drug's effect was in gray zone and need to pause and think, I would change the label of a=0 in Figure 3 to reflect that fact. I also think that we don't need this extra diagram since it is quite straightforward. ) \xx

%\rr In an ASIED, probability thresholds $\xi_1$ and $\xi_2$, sample size to perform the first interim and the second interim $n_1$ and $n_2$ are design parameters that need to be determined by operating characteristics through simulations.\xx 

\section{A Bayesian Subgroup Identification  Model}
\label{sec:model}
Recall that in Section  \ref{sec:design} we assume there exists a subgroup identification model,  which can identify the subgroup with enhanced treatment effect based on patients' biomarker profiles  and responses  at each interim analysis. In our implementation we introduce a  flexible Bayesian model as  the core search algorithm that can handle biomarkers of varying forms and different types of  response  outcomes, 
building upon the methods proposed in \cite{xu2014subgroup} and \cite{guo2017subgroup}. 
%Xu et al. (2014) developed SUBA, a Bayesian subgroup-based adaptive design, to allocate the patients to their superior treatments using a random partition model that splits the biomarker space by the observed biomarker?s median value, which generally is not the optimal cutoff for a predictive biomarker. Guo et al. (2017) extended SUBA (i.e., SCUBA) by allowing the biomarker space to be split using hyperplanes that construct linear boundaries, providing a more flexible partition model. 
The proposed model will be easy to implement and accurately identify the subgroup with enhanced treatment effect when the sample size $N$ is small. 
However, any alternative subgroup identification model that learns about enhanced treatment effects by including treatment by covariate interactions, could be used. For example, when the sample size is large, machine learning-based methods such as  random forest \citep{foster2011subgroup} or more flexible Bayesian methods such as Bayesian additive regression trees (BART) \citep{chipman2010bart} could be incorporated into the decision-making framework of the proposed ASIED. 

Assume we have a maximum sample size of $N$ patients that are indexed by $i = 1, \ldots, N$, and suppose there are $T$ candidate treatments  indexed by $t = 1, \ldots, T$. 
%In the motivating AD trial, $T=2$. 
Let $z_i = t$ denote that patient $i$ is assigned to treatment $t$. Assume that we have $K$ biomarkers that are potential predictive biomarkers 
 identified from the investigational drug's mechanism of action or disease clinical presentation. We assume that a biomarker $k$ can be binary, ordinal, categorical, %with $b \in \{1, \ldots L_b\}$ where $L_b$ is the set of all possible labels for biomarker $b$, 
or continuous, where $k=1, \dots, K$. %with $b \in \mathbb{R}$ which is then standardized so that $b$ is in the range $[-1,1]$. 
Denote $\xb_i = (x_{i1}, \dots , x_{iK})'$ and $y_i$ to be the biomarker profile and the response outcome of the $i^{th}$ patient, respectively.

%. Let $y_i$ denote patient $i$'s outcome.

%We make explicit the assumption that in a clinical trial, there are subgroups of patients across which a single treatment may have different responses. We do not know a priori what these subgroups are, and must estimate them from response and biomarker measurement data. We propose an adaptive estimation scheme, wherein we use previously treated patients' data to make future predictions. Our model is a random partition model.
Let $\Omega$ denote the biomarker space. We say that a partition is a family of subsets $\Pi = \lbrace S_1, \cdots, S_m, \cdots, S_M\rbrace$, where the $S_m$'s are mutually disjoint and their union is $\Omega$.  Here the number $M$ of subsets is random.  The partition on the biomarker space  induces a partition of the patients. If $\xb_i \in S_m$, we say patient 
$i$ with biomarker profile $\xb_i$ belongs to subgroup $m$. We will construct a prior probability measure for $\Pi$ in the next section. Below we consider the sampling model of $y_i$ of different types conditional on $\xb_i$ and $\Pi$, using binary outcomes and continuous outcomes as examples. It can be easily extended to other types of outcomes, such as categorical or survival outcomes. 
Let $\Theta$ denote the parameters in the sampling model. Denote $\bm{Y}_n = (y_1, \ldots ,y_n)$, $\bm{X}_n = \{\xb_i\}^n_{i=1}$, and $\bm{Z}_n = (z_1, \ldots , z_n)$. 

\noindent{\underline{Binary outcomes}}. Let $y_i \in \{0, 1\}$ and $\theta_{t, m}$ be the response rate of patients in subgroup $m$ under treatment $t$.  In this case, 
$\Theta = \{\theta_{t, m}\}_{t=1,m=1}^{T, \ \ M}$. 
We assume
 $$
p(y_i = 1 \mid z_i =t, \Pi, \xb_i \in S_m) = \theta_{t, m}.$$
The likelihood function is simply the product of $n$ Bernoulli  probability mass functions. We assign the prior $\theta_{t,m} \mid \Pi \iid Beta(a_t, b_t)$, where $Beta(a_t, b_t)$ denotes a beta distribution with mean $a_t/(a_t+b_t)$.

%{\underline{Categorical outcomes}}.
%Let $y_i \in \{1, 2, \dots, C\}$, where $C$ is the number of outcome categories. Denote $\theta_{c, t, m}$ to be the response rate of patients in subgroup $m$ under treatment $t$ for outcome category $c$. We assume
%$$
%p(y_i = c \mid z_i =t, \Pi, \xb_i \in S_m) = \theta_{c, t, m},
%$$
%where $\sum_{c=1}^C \theta_{c,t,m}=1$. We assign the conjugate prior $(\theta_{1, t, m}, \dots, \theta_{C, t, m})\mid \Pi\iid \mathrm{Dirichlet}(a_{t1}, \dots, a_{tC})$. 

\noindent{\underline{Continuous outcomes}}. 
Let $y_i \in R$ and $\theta_{t, m}$ be the mean response of patients in subgroup $m$ under treatment $t$.
We assume
$$
y_i \mid z_i =t, \Pi, \xb_i \in S_m \sim N(\theta_{t, m}, \sigma^2).
$$
The likelihood can be written as follows:
%Let $y_{i; t,m}$ denotes the $i^{th}$ patient's response out of the subset of patients that belong to subgroup $S_m$ and receive treatment $t$ and $\theta_{t,m}$ be the response of patients in subgroup $S_m$ under treatment $t$. We have that, $y_{i;t,m} = \theta_{t, m} + \epsilon_{i}$ where $\epsilon_{i} \sim N(0, \sigma^2)$. We let $\Theta = \{\theta_{t, m}, \sigma^2\}$. The sampling model can then be written as the following, 
\begin{equation}
p(\bm{Y}_n \mid \bm{X}_n, \bm{Z}_n, \Theta, \Pi) = \prod_{t=1}^{T}\prod_{m=1}^{M}\prod_{\{i: z_i=t, \xb_i\in S_m\}}(2\pi\,\sigma^2)^{-1/2} \exp\{-\frac{1}{2\sigma^2}(y_{i} - \theta_{t, m})^2\}.
\end{equation}

We assign the conjugate prior $p(\theta_{t, m}, \sigma^2) = p(\theta_{t, m} | \sigma^2)p(\sigma^2)$ with $\theta_{t, m} | \sigma^2 \sim  N(\theta_0, \frac{\sigma^2}{\kappa_0})$ and $\sigma^2 \sim IG(\frac{\nu_0}{2}, \frac{SS_0^2}{2})$, where $SS^2_0 = \nu_0 \sigma^2_0$.

%{\underline{Survival outcomes}}.
%We assume that $\log(y_{i}) = \xb'_{i}\,\bm{\beta}_{t, m} + \epsilon_{i}$, where $\bm{\beta}_{t, m}$ denotes the regression coefficient for patients in subgroup $m$ under treatment $t$ and $\epsilon_{i} \sim N(0, \sigma^2)$.  We assign the priors $\bm{\beta}_{t,m} \sim N(\mu_0, \Sigma_0)$ and $\sigma^2 = IG(a_0, b_0)$.
%\begin{equation}
%p(y_{i} | x_{i}, \bm{\beta}_{t, m}, \sigma) = \frac{1}{\sigma \sqrt{(2\pi)}} \exp{(-\frac{(y_{i} - x_{i}^T\,\bm{\beta}_{t, m})^{T}\,(y_{i} - x_{i}^T\,\bm{\beta}_{t, m})}{2\sigma^{2}})}.
%\end{equation}

%We assign the priors $\bm{\beta}_{t,m} \sim N(\mu_0, \Sigma_0)$ and $\sigma^2 = IG(a_0, b_0)$.

The joint model can be written as follows,
\begin{eqnarray}
p(\bm{Y}_n, \Theta, \Pi \mid \bm{X}_n, \bm{Z}_n) \propto p(\bm{Y}_n \mid \bm{X}_n, \bm{Z}_n, \Theta, \Pi)\, p(\Theta \mid \bm{X}_n, \bm{Z}_n, \Pi)\, p(\Pi \mid \bc)p(\bc), 
\label{eq:model}
\end{eqnarray}
where $\bc$ denotes the parameters in the prior model that describes the random partition $\Pi$. 
We have introduced the sampling model and the priors for $\Theta$. In the next section, we will discuss the prior of $\Pi$ and $\bc$. 

\subsection{Prior of Partition}
We propose a tree-type random partition on the biomarker space $\Omega$ to  define random  biomarker subgroups.  
We build partitions via a tree of recursive splits: each node of the tree represents a subset of $\Omega$. The final leaves of the tree are the partitioning sets $\{S_m\}_{m=1}^M$.  
At each node the tree is either pruned or the corresponding subset is further split into two siblings. In the latter case, the two siblings are defined by a plane orthogonal to a randomly selected axes of $\Omega$, say the axis of the $k$-th biomarker. 
%This plane intersects and splits the ancestor node. 
In other words, through a sequence of splits, each of which selects a biomarker $k$ first and then splits the space of $x_k$ into two subspaces, we generate a partition set of $\Omega$ as the collection of the resulting subsets. For the motivating AD trial, we limit the partition to at most four biomarker subgroups due to the small sample size, and hence there are no more than two rounds of random splits in the random partition. This constraint is imposed to  avoid resulting in subgroups with too few patients. 

\begin{figure}[h]
\centering
\includegraphics[scale=0.5]{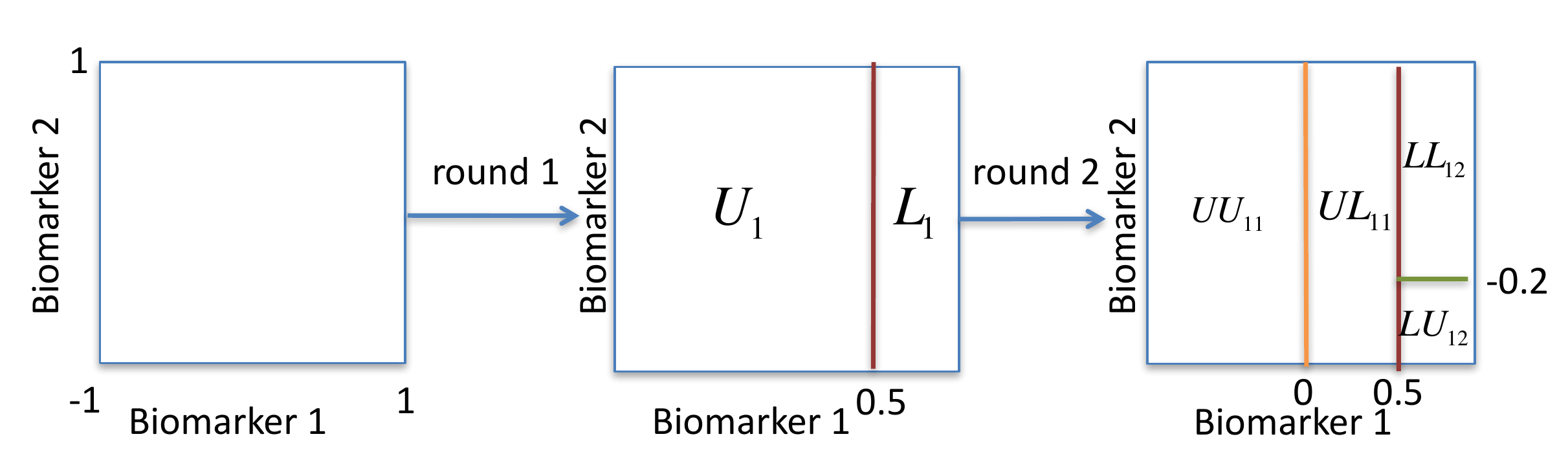}
\caption{An illustration of $\Pi$. The example shows that with two rounds of split, the initial space of two biomarkers is partitioned into four subsets $ \{UU_{11},UL_{11},LU_{12}, LL_{12}\}$.}
\label{fig:partition}
\end{figure}

Figure \ref{fig:partition}) illustrates the procedure of random partition using a simple example with two rounds of splits and two continuous biomarkers on $[-1, 1]^2$.  
In each round, for each of the current subsets, we split along a biomarker $k$ with probability $\nu_k$ or choose not to split with probability $\nu_0$, $\sum_{k=0}^K\nu_k=1$. If an ancestor subset $S$ is split into two subsets by the $k^{th}$ biomarker, then the resulting subsets are $\{i: x_{ik}\leq c_k(S)\}$ and $\{i: x_{ik}> c_k(S)\}$, where $c_k(S)$ is the threshold by which the subset is being split. We will discuss the prior of $c_k(S)$ later. For example, in Figure \ref{fig:partition}, biomarker 1 is chosen in the first round and the patients are split into $U_1=\{i: x_{i1}\leq 0.5\}$ and $L_1=\{i: x_{i1}> 0.5\}$. Here the subindex $_1$ denotes that biomarker 1 is chosen to split; $U$ and $L$ denote that the measurements are smaller and larger than the threshold, respectively.  In round 2, we split the subgroup $U_1$ into two biomarker subgroups $UU_{11}$ and $UL_{11}$ by choosing biomarker 1 with threshold 0 and split the subgroup $L_1$ into two biomarker subgroups $LU_{12}$ and $LL_{12}$ by choosing biomarker 2 with threshold -0.2. 
%The split is conducted based on the conditional median of biomarkers 1 and 2 respectively, i.e. the median of biomarker 1 within the subgroup U1 and the median of biomarker 2 within the subgroup L1. 
Note that the ordering of letters U and L are matched with the ordering of the biomarker index. Therefore, at the end, we have the partition $\Pi = \{UU_{11},UL_{11},LU_{12}, LL_{12}\}$, which corresponds to four biomarker subgroups.

For a given subset $S$, if biomarker $k$  is selected with probability $\nu_k$ from a set of available biomarkers to split at threshold $c_k(S)$, the prior of  $c_k(S)$  depends on the type (continuous, binary, categorical, or ordinal) of biomarker $k$. 
By ``available", we mean that the split rules would not lead to empty subgroup. For example, if a binary biomarker was used in one round of split, then it would no longer be available for splitting at nodes below it.  We describe the split rules for determining a partition of the biomarker space for various types of biomarkers, including continuous, binary, categorical, and ordinal biomarkers as follows. 

\begin{itemize}
\item If biomarker $k$ is binary, the split will be deterministic and we denote $U_k=\{i: x_{ik} = 0\}$ and $L_k=\{i: x_{ik} = 1\}$.  Therefore $p(c_k) = 1$. 

\item If biomarker $k$ is continuous, denote $U_k=\{i: x_{ik} \leq c_k\}$ and $L_k=\{i: x_{ik} > c_k\}$. We assume $p(c_k)=\mathrm{Uniform}( \min \{ x_{ik} \}_{i=1}^N,  \max \{ x_{ik} \}_{i=1}^N)$. 

\item If biomarker $k$ is ordinal, let $V_k$ denote the number of labels that biomarker $k$ has. Let $c_k$ denote the endpoint of the left partition, e.g., if $V_k = 5$ and $c_k = 3$, the left partition is $\lbrace 1, 2, 3\rbrace$ and the right partition is $\lbrace 4, 5\rbrace$. In this way we denote $U_k=\{i: x_{ik} \leq c_k\}$ and $L_k=\{i: x_{ik} > c_k\}$. Moreover, if $c_k = V_k$, it is equivalent to not splitting, which has been considered with probability $\nu_0$. 
Therefore, $p(c_k) = \frac{1}{V_k - 1}$.

\item If biomarker $k$ is categorical, let $V_k$ denote the number of categories corresponding to biomarker $k$. Let $c_{k}$ denote the elements in one subset $U_k$. The remaining elements are stored in the other subset  $L_k$. The $c_k$ are elements of the powerset of $\lbrace 1, 2, \cdots, V_k\rbrace$ without the empty-set or the full set. There are hence $2^{V_k} - 2$ options for $c_k$. Note that the choice of $c_k$ is symmetric: we may flip $c_k$ and its complement, leading to the same partition. Thus, 
%if $k = K+1,\, p(c_k) = 1$. Otherwise, 
$p(c_k) = \frac{2}{2^{V_k} - 2}$.
\end{itemize}

In the Supplement A, we describe the detailed split rules using two rounds of splits as an example by taking into account various types of biomarkers. 

%Compared to SUBA \citep{xu2014subgroup}  and SCUBA \citep{guo2017subgroup}, our contributions are two folds. First, SUBA and SCUBA only deal with continuous biomarkers, while the proposed method considers multiple types of biomarkers. Second, SUBA splits the selected biomarker by the observed biomarker?s median value
% as the cutoff, which is not always the best choice. SCUBA extends SUBA by allowing to split using hyperplanes that construct linear boundaries

\subsection{Posterior Inference}
Markov chain Monte Carlo (MCMC) method is used to obtain the posterior samples of the parameters based on the joint model \eqref{eq:model}. 
Each iteration of the MCMC simulations consists of the following transition probabilities. 
Sampling $\theta_{t,m}$ is straightforward due to the use of conjugate priors. For example in the binary outcome case, we can easily compute that $p(\theta_{t,m}\mid \bY_n, \Pi, \bZ_n) \sim \mathrm{beta}(n_{tm1}+a, n_{tm0}+b)$, where $n_{tmy}=\sum_i I(\xb_i\in S_m, z_i=t, y_i=y)$, $y=0, 1$. In the joint model \eqref{eq:model}, $p(\Pi\mid \bc)=1$ since $\bc$ decides the partition $\Pi$ deterministically. Sampling the biomarkers  $k$'s that split the patient subsets might change the number of subgroups $M$, leading to the change of the dimension of $\bc$, since we allow to choose not to split. Hence we make use of a two-step Metropolis-Hasting sampler to first update the thresholds $c_k$'s conditional on the selected biomarkers, then update both biomarkers $k$'s and their thresholds $c_k$'s  together to ensure fast convergence. We defer the detailed MCMC derivations for all types of outcomes in the Supplement B. 

\subsection{Subgroup Identification}
\label{sec:subgroupidenfication}

Since a random distribution is proposed as a prior on the partition, summarizing a distribution over the random partition and then identifying the subgroups with enhanced treatment effects become challenging. Reporting subgroups with enhanced treatment effects hinges on the discovery of regions in the biomarker space in which one treatment outperforms the others. 
Assume $n$ patients have been treated and their responses have been  obtained at  the interim analysis. Denote $\mathcal{D}_n=\{\bm{Y}_n, \bm{X}_n, \bm{Z}_n \}$. We define an equally spaced grid of $D_k$  values $\{x_{k1}, \dots, x_{kD_k} \}$  for biomarker $k$  in the biomarker space, 
then  take the Cartesian product of the grids across all $K$ biomarkers to obtain a $K$-dimensional grid $\tilde{\xb}_d$ of size $\prod_kD_k$ points. 
  For example, if biomarker $k$ is continuous on [-1, 1], we can choose $D_k=20$ equally spaced points on [-1, 1]; if biomarker $k$ is binary, then $D_k=2$. Each grid point $d$ represents a possible biomarker profile.  For patient $i$ with biomarker profile $\tilde{\xb}_d$, we can compute the posterior predictive distribution of $y_i$ by 
 $$p(y_i\mid z_i=t, \xb_i=\tilde{\xb}_d, \mathcal{D}_n)=\int p(y_i\mid \Theta, z_i=t, \xb_i=\tilde{\xb}_d)p(\Theta\mid \mathcal{D}_n)d\Theta.$$ 
 In the MCMC samples, the $b^{th}$ iteration after burn-in generates a posterior sample $\{ \Theta^{(b)}, \Pi^{(b)}, \bc^{(b)}\}$, which defines a partition set $\Pi^{(b)}=\{ S^{(b)}_1, \dots, S^{(b)}_{M^{(b)}} \}$ and their corresponding response parameters. We can easily calculate the posterior mean response of the patient with biomarker profile $\tilde{\xb}_d$ at iteration $b$ as  $\theta^{(b)}_{t, m}$
 %$\hat{\theta}_{t, d} = \frac{1}{B}\sum_b\theta^{(b)}_{t, m}$ 
 if $z_i=t$ and $\tilde{\xb}_d\in S^{(b)}_m$, here $B$ is the number of saved MCMC iterations after burn-in.

In the motivating AD trial, we assume that $t=1$ represents placebo and $t=2$ represents the investigational drug. %Denote the difference in posterior mean responses $q_d = \hat{\theta}_{2, d} - \hat{\theta}_{1, d}$, then 
The posterior estimated treatment effect for a given subgroup $\Delta$ at iteration $b$ can be represented by 
%$$\delta_{\Delta} = \frac{1}{n_{\Delta}}\sum_{d: \tilde{\xb}_d\in \Delta} \frac{1}{B}\sum_bI\big[(\hat{\theta}^{(b)}_{2, d} - \hat{\theta}^{(b)}_{1, d})>\mathrm{LRV}\big],$$
$$\delta^{(b)}_{\Delta} = \frac{1}{n_{\Delta}}\sum_{d: \tilde{\xb}_d\in \Delta} (\hat{\theta}^{(b)}_{2, d} - \hat{\theta}^{(b)}_{1, d}),$$
where $n_{\Delta}$ denotes the number of grid points that fall in the subgroup $\Delta$. Denote the posterior probability of treatment effect being larger than LRV for a given subgroup $\Delta$ by $Pr(\delta_{\Delta}\geq \mathrm{LRV})$. We can easily compute 
$$Pr(\delta_{\Delta}\geq \mathrm{LRV}) = \frac{1}{B}\sum_bI\big[\delta^{(b)}_{\Delta}>\mathrm{LRV}\big], $$ 
which can be then used for the Go/Stop/Gray zone decision-making described in Section \ref{sec:design}. 
If the goal is to identify the subgroup of patients whose posterior probability of treatment effect being larger than LRV is bigger than a threshold $\xi$, we can represent such a subgroup as $\{\Delta: Pr(\delta_{\Delta}\geq \mathrm{LRV}) >\xi\}$. 

\subsection{Simulation Studies on Subgroup Identification}

We conducted simulation studies to evaluate the proposed Bayesian random partition (BayRP) model 
 on subgroup identification. 
We designed simulation scenarios based on the motivating AD trial as described in Section \ref{sec:motivate} and assumed that $K=4$ baseline biomarkers were available for each patient and $p(\nu_k) = 1/5$, $k=0, 1, \dots, 4$, indicating a uniform prior on the biomarker selection. The priors on the parameters in $\bc$ were introduced in supplement A. 
We considered four scenarios and simulated 100 trials for each scenario.  In the motivating AD trial, the first interim analysis will be conducted when 100 subjects have finished Week 12 visit of the study. Therefore, 
the sample size $n=100$ was set for each scenario to test if BayRP can accurately identify subgroups with enhanced treatment effects. 
% since in the motivating AD trial, \rr the first interim analysis will be conducted when 100 subjects have finished Week 12 visit of the study. 
 With equal randomization to the placebo or the investigational drug, this would give us approximately 50 per group for the interim analysis. 

In the first three scenarios, we assumed all the biomarkers were continuous and generated $x_{ik}$ from $\mathrm{Uniform}(-1, 1)$, $i = 1,  \dots, n$ and $k=1, \dots, 4$. 
In \underline{scenario 1}, we assumed only the first biomarker was related to the response and $y_i=0.75 + 0.25I(z_i=2) + 3 I(x_{i1}>-0.4)I(z_i=2) + \epsilon_i$. In \underline{scenario 2},  the first two biomarkers were related to the response and 
the outcomes $y_i$'s were generated from $y_i=0.75 + 0.25I(z_i=2) + 3 I(x_{i1}< 0.3, x_{i2}>-0.4)I(z_i=2) + \epsilon_i$. In \underline{scenario 3}, we assumed $y_i=0.75 + 0.25I(z_i=2) + 1.5 I(x_{i1}> 0.4)I(z_i=2) + \epsilon_i$. In \underline{scenario 4}, we considered three continuous biomarkers generated from $\mathrm{Uniform}(-1, 1)$ and one binary biomarker generated from $\mathrm{Bernoulli}(0.5)$. 
The response $y_i$'s were generated from $y_i=0.75 + 0.25I(z_i=2) + 3.5 I(x_{i1}=1, x_{i2}>-0.4)I(z_i=2) + \epsilon_i$. 
Here $\epsilon_i\sim N(0, 1)$. 

\begin{figure}[h!]
\centering
\begin{tabular}{cccc}
Truth &  BayRP& LR &GUIDE\\
\includegraphics[width=.23\textwidth]{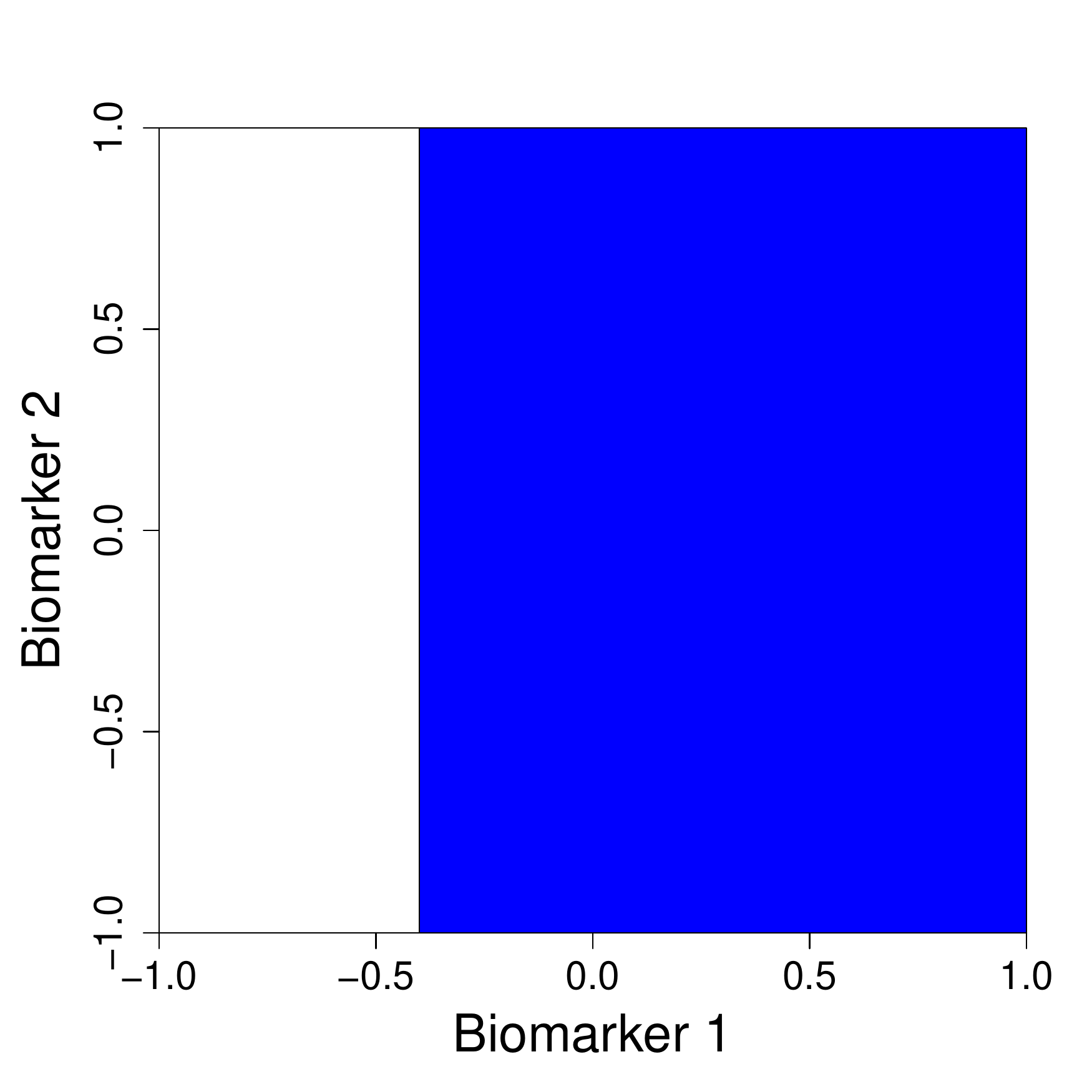} & \includegraphics[width=.23\textwidth]{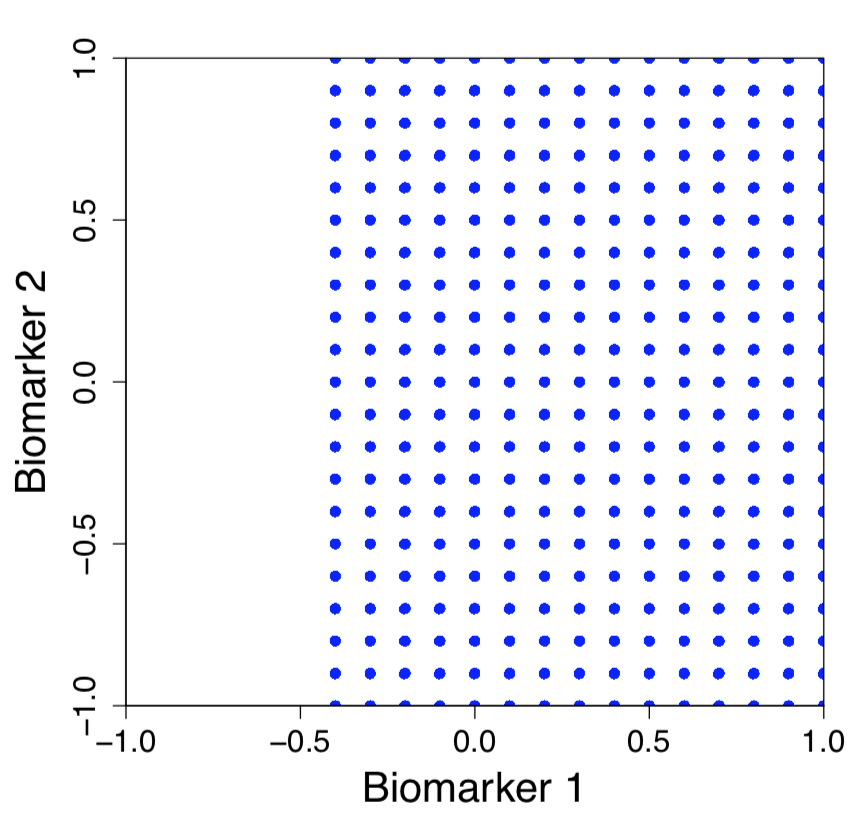}&\includegraphics[width=.23\textwidth]{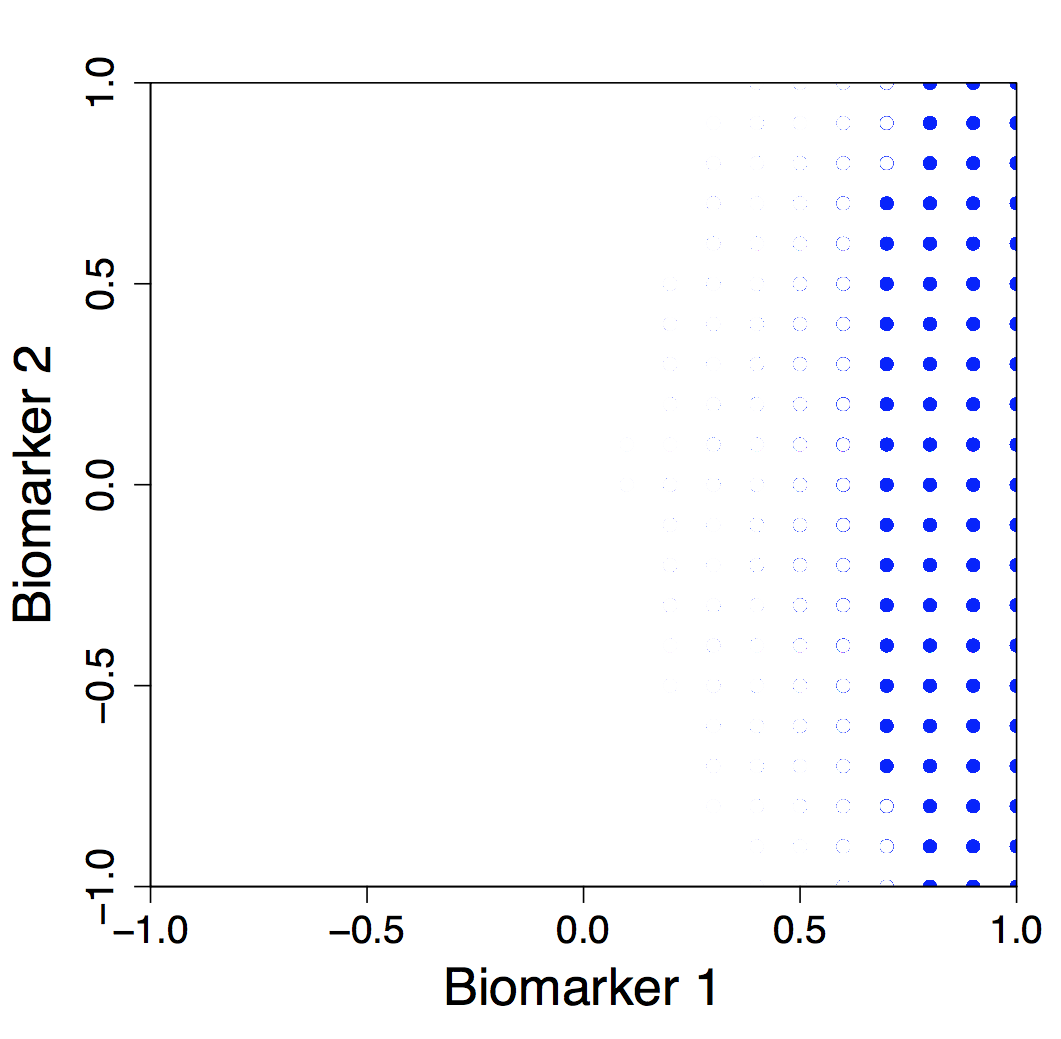} & \includegraphics[width=.23\textwidth]{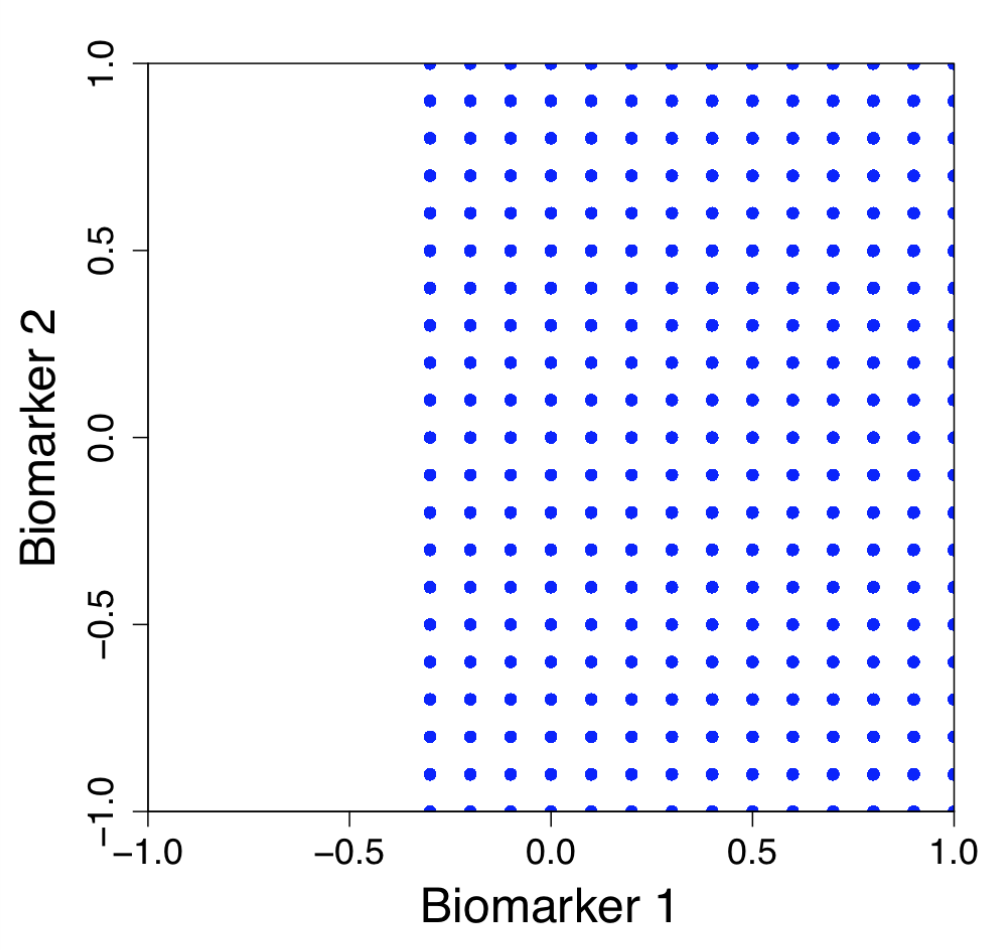} \\
 Scenario 1 &  Scenario 1& Scenario 1& Scenario 1\\
\includegraphics[width=.23\textwidth]{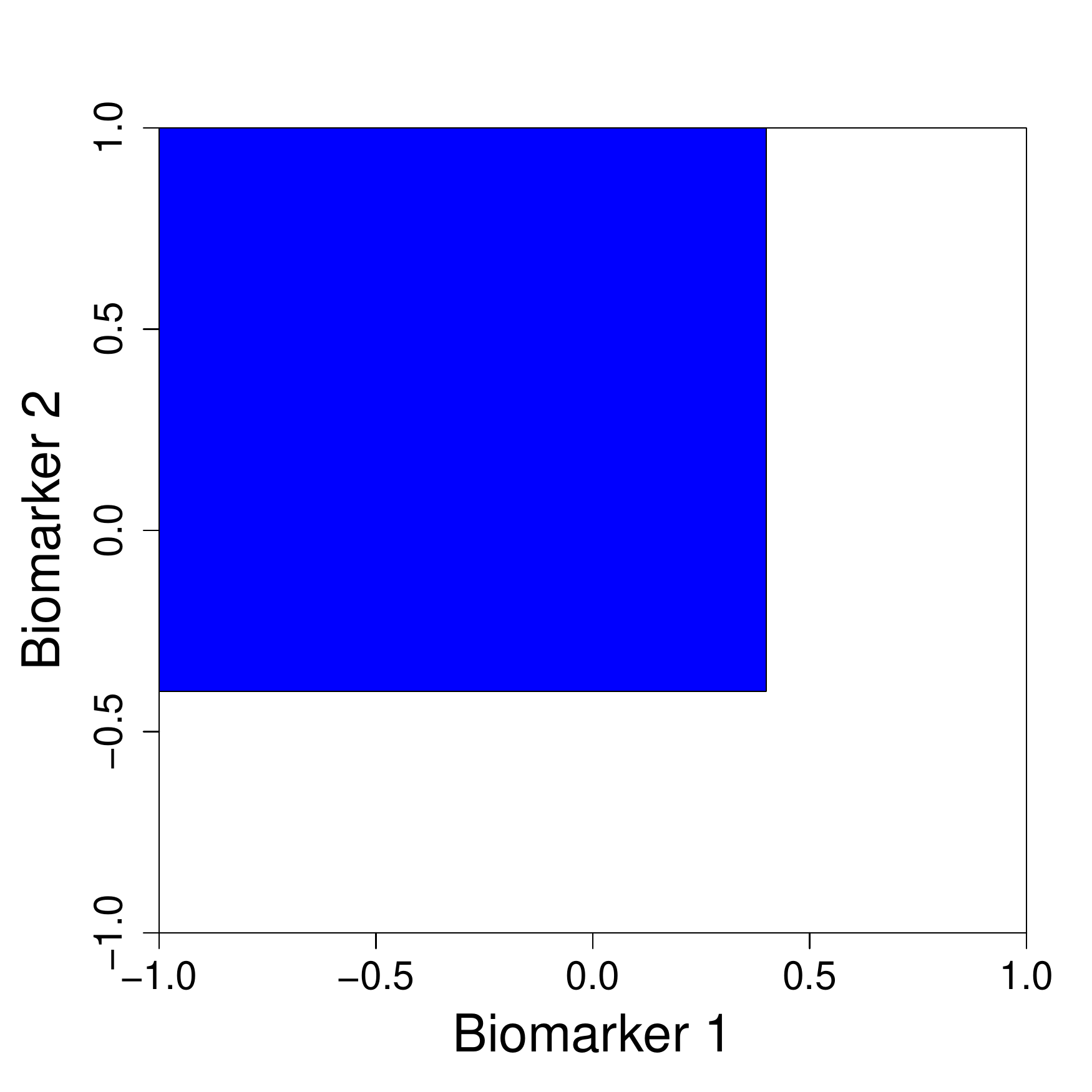} & \includegraphics[width=.23\textwidth]{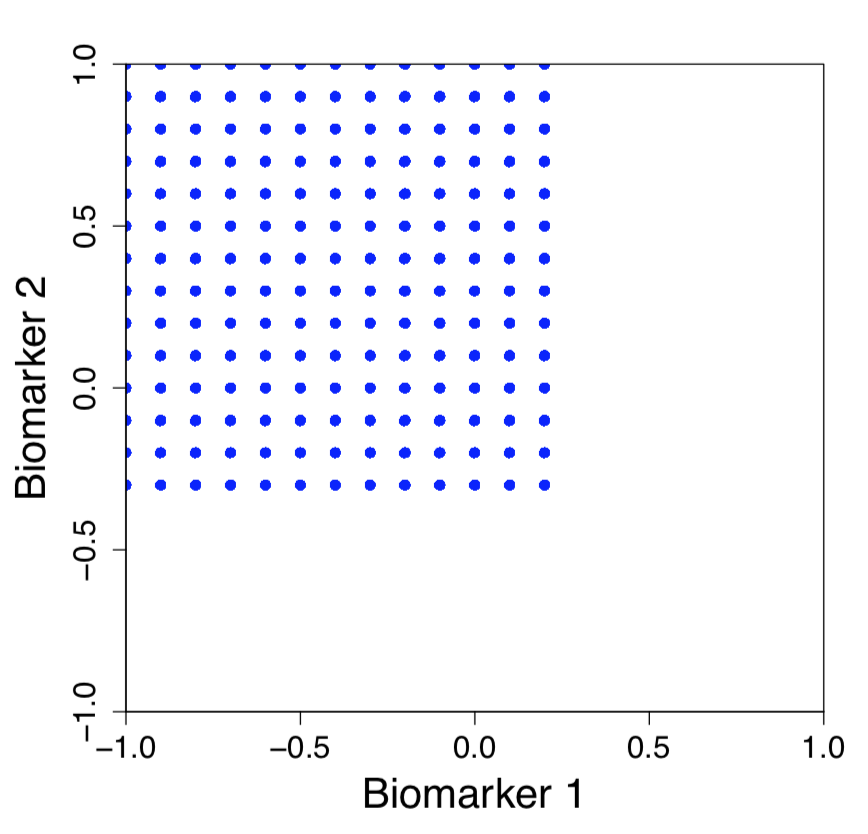}&\includegraphics[width=.23\textwidth]{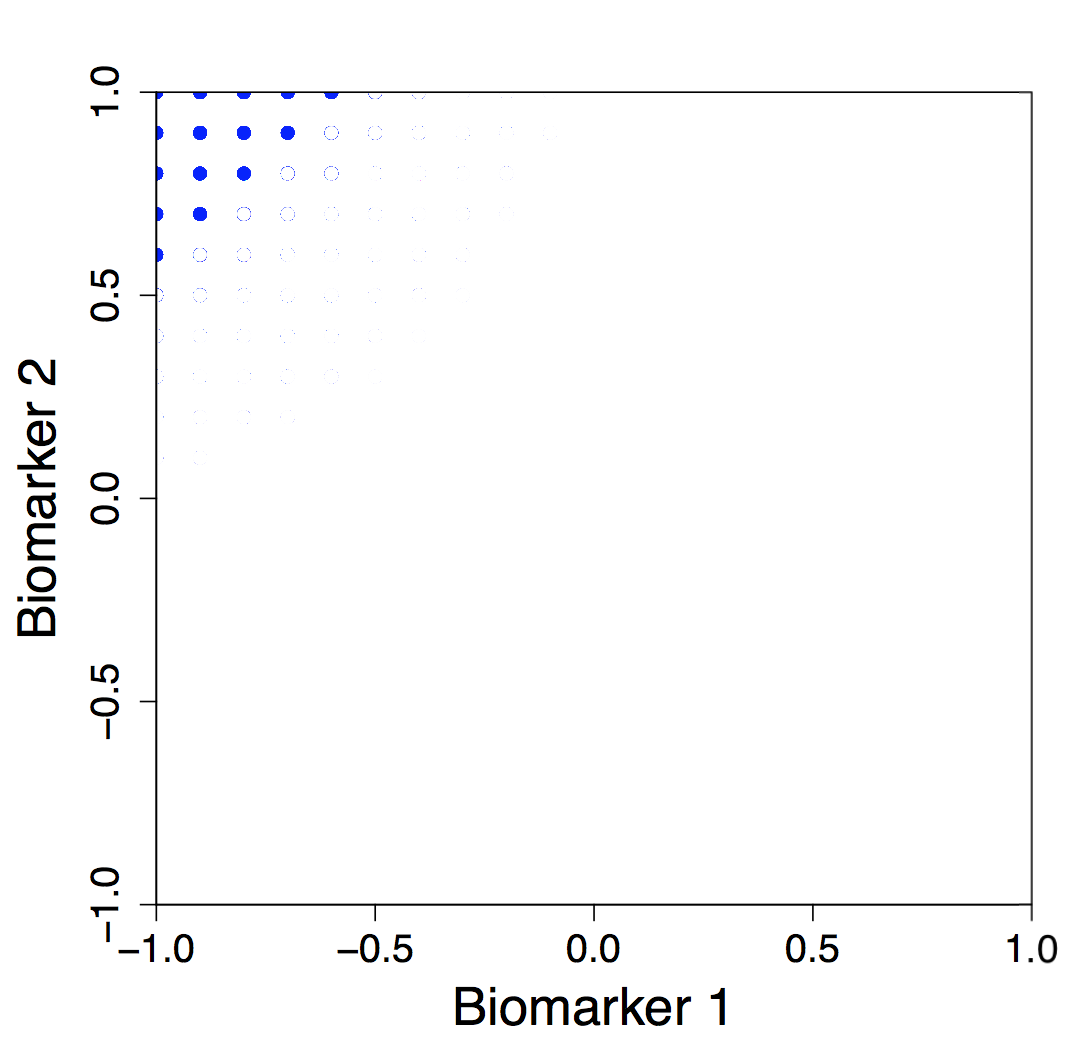} & \includegraphics[width=.23\textwidth]{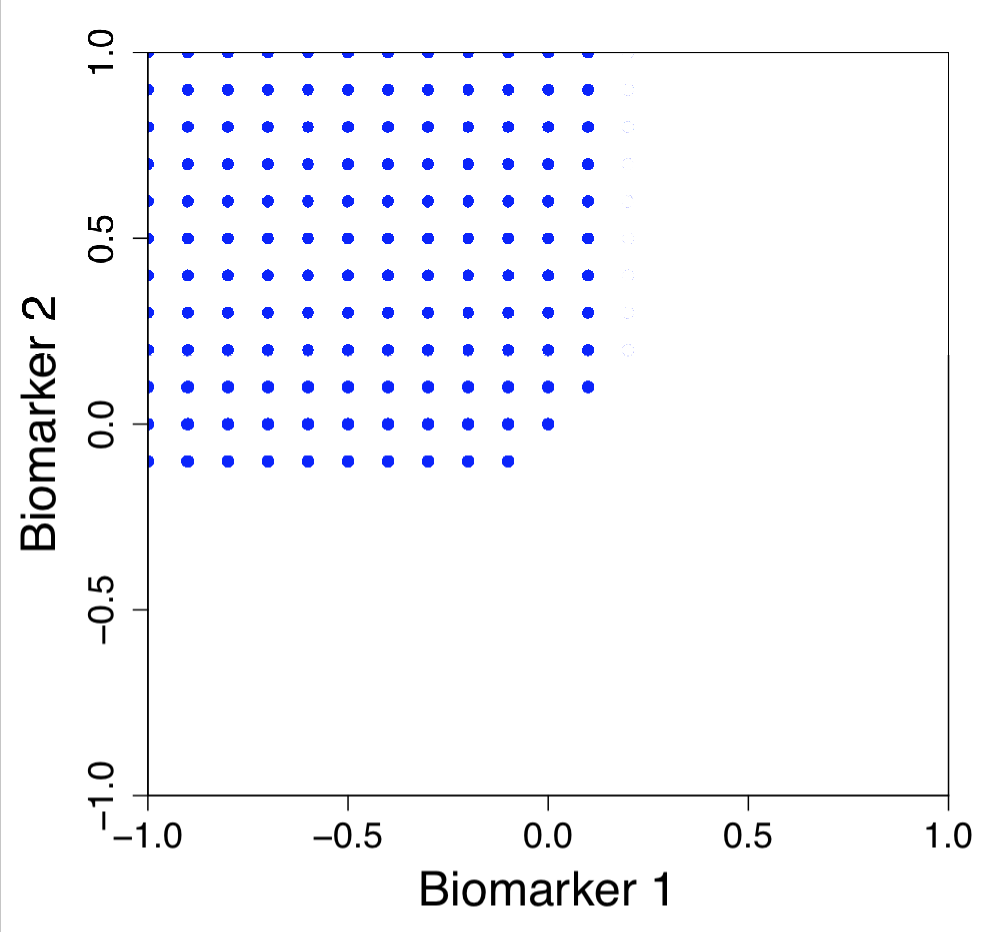} \\
 Scenario 2 &  Scenario 2& Scenario 2& Scenario 2\\
\includegraphics[width=.23\textwidth]{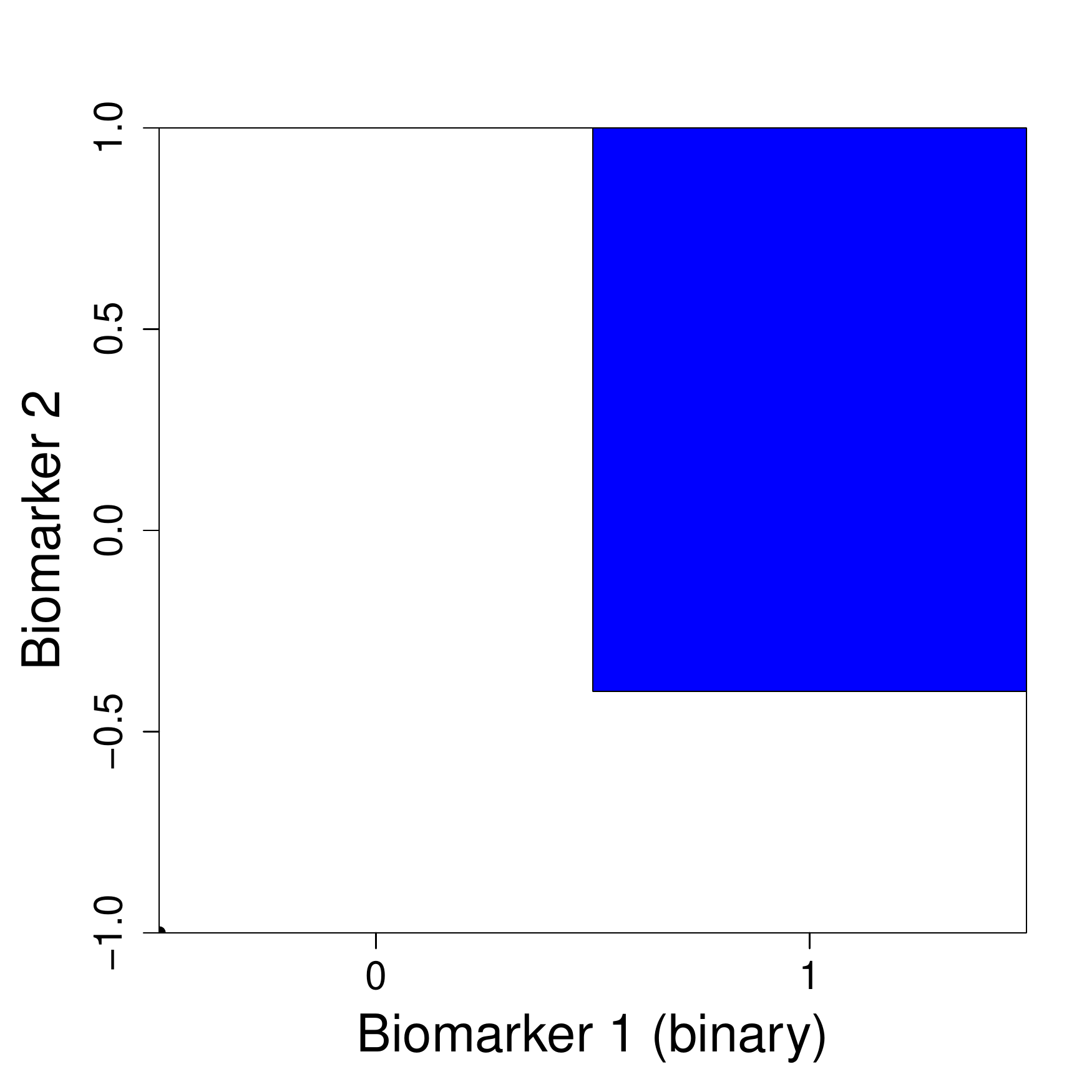} & \includegraphics[width=.23\textwidth]{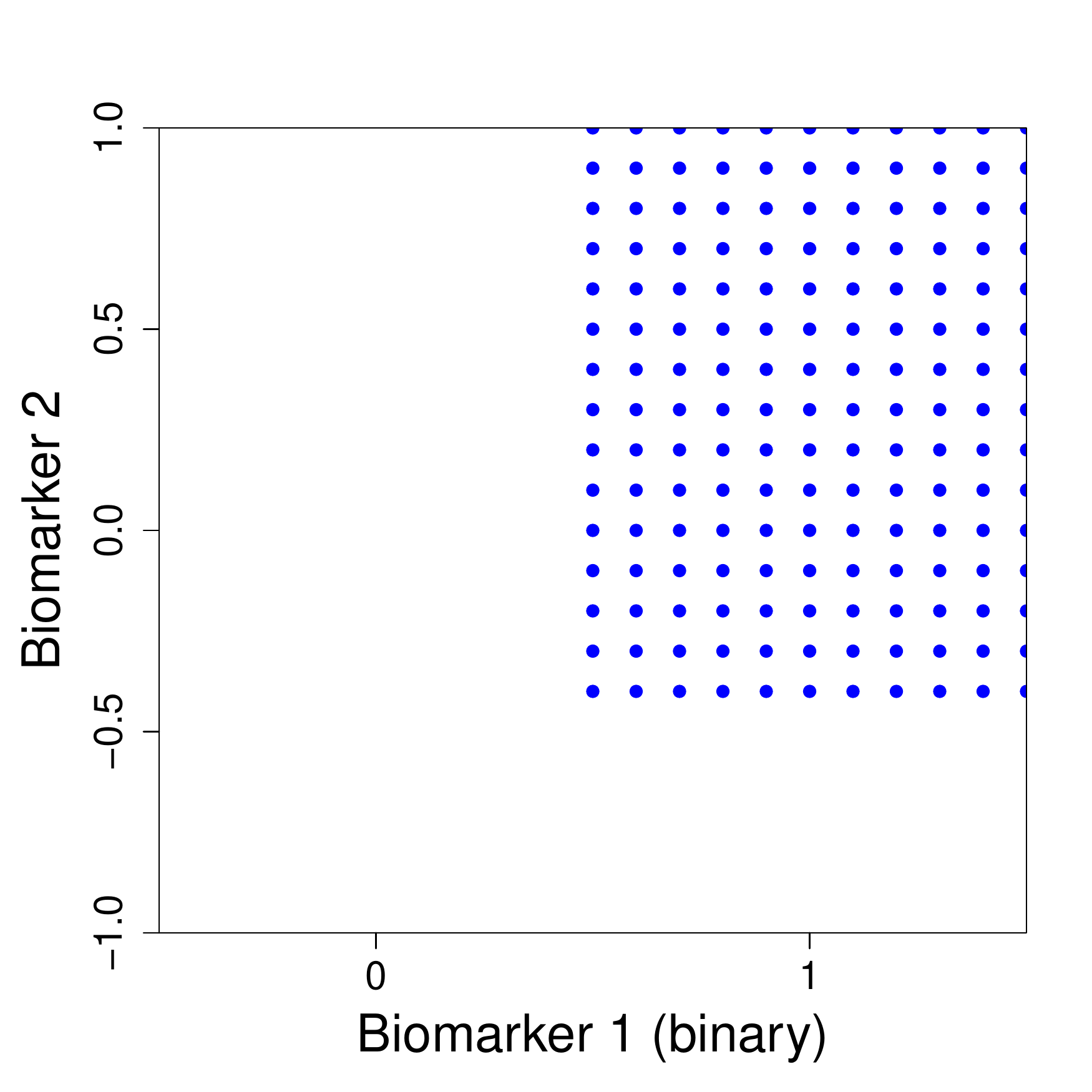}&\includegraphics[width=.23\textwidth]{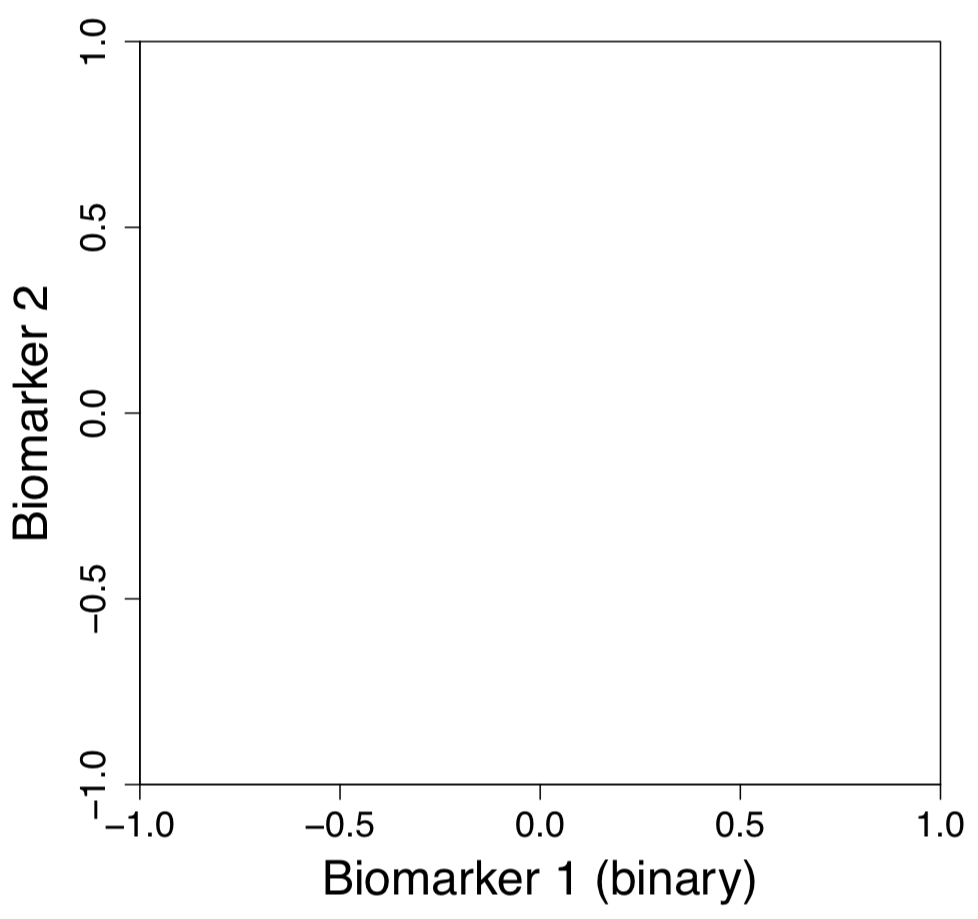}&\includegraphics[width=.23\textwidth]{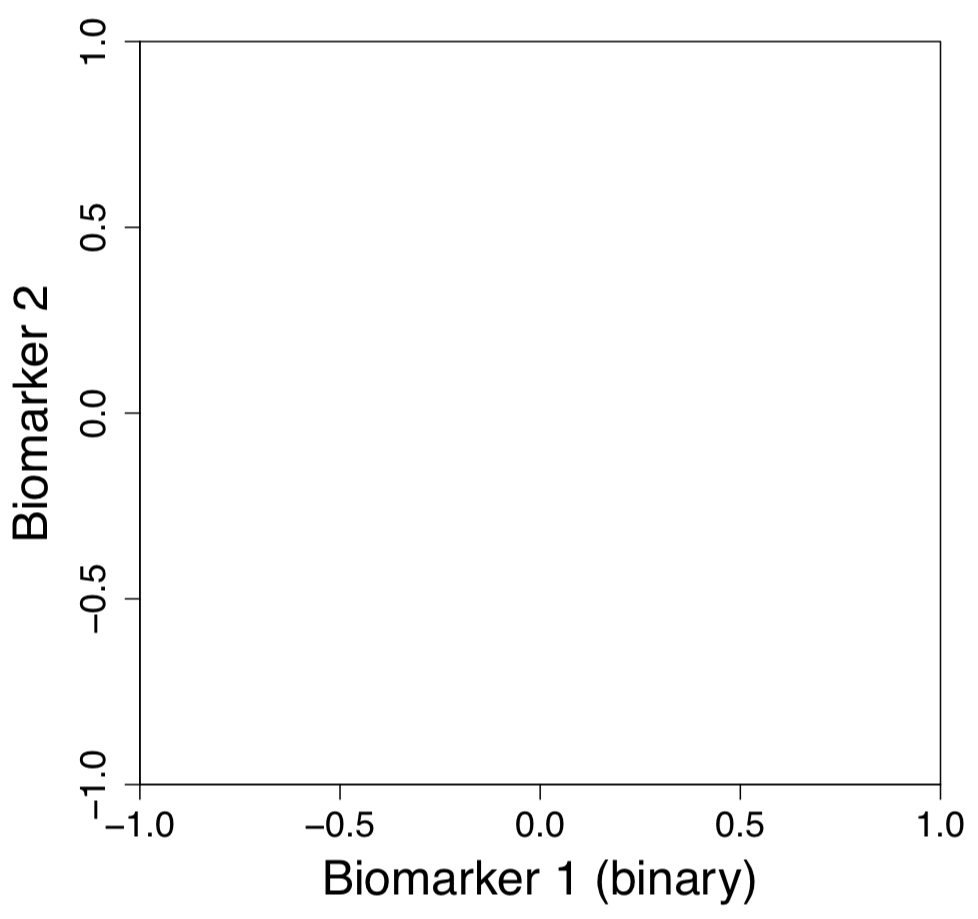} \\
 Scenario 4 & Scenario 4& Scenario 4& Scenario 4\\
\end{tabular}
\caption{From left to right: true effective subgroups in blue for scenarios 1, 2, and 4; the posterior estimated subgroups represented by grid points in blue under the BayRP, LR, and GUIDE, respectively. }
\label{fig:subgroup}
\end{figure}

Define the true effective subgroup as $S^o = \big\{i: [E(y_i\mid z_i=2, \xb_i)-E(y_i\mid z_i=1, \xb_i)] > \mathrm{LRV}\big\}$. In the motivating AD trial, LRV=2.37 was used.  The left column of Figure \ref{fig:subgroup} shows the simulated true effective subgroups in blue color for scenarios 1, 2, and 4. Note that we have four biomarkers, but only biomarkers 1 and/or 2 are predictive to responses. Therefore, we plot the true simulated effective subgroup versus biomarkers 1 and 2. 
Scenario 3 is not shown since there is no   effective subgroup. 

We first report the effective subgroup estimated by the proposed Bayesian random partition (BayRP) model. For each simulated trial $h$, $h=1, \dots, 100$, 
we denote $\delta_d^{h, (b)}=\hat{\theta}^{h, (b)}_{2,d}-\hat{\theta}^{h, (b)}_{1,d}$ to be the posterior estimated treatment effect of grid $d$ with biomarker $\tilde{\xb}_d$ at iteration $b$ in trial $h$. The collection over all the MCMC iterations can be used to report the effective subgroup. Denote $\hat{\delta}_d^{h}$ to be the probability that the treatment effect of the grid $d$ with biomarker $\tilde{\xb}_d$ is larger than LRV  in trial $h$. 
Given a desired confidence $\xi$, we determine that the grid $d$ with biomarker $\tilde{\xb}_d$ belongs to the effective subgroup in trial $h$ if 
\begin{eqnarray}
\hat{\delta}_d^{h}=\frac{1}{B}\sum_bI(\delta_d^{h, (b)}\geq \mathrm{LRV})>\xi.
\label{eq:sub}
\end{eqnarray}
Here we set $\xi=0.9$.  
Then we define the estimated effective subgroup $\hat{\Delta}_{\mathrm{BayRP}}$ under BayRP  as the set of subjects whose posterior probability of the treatment effect being larger than LRV are bigger than 0.9, which can be computed as 
\begin{eqnarray}
\hat{\Delta}_{\mathrm{BayRP}} = \Big\{d: \frac{1}{100}\sum_{h=1}^{100}I(\hat{\delta}_d^{h}> \xi)>0.9\Big\}.
\label{eq:ASIED}
\end{eqnarray}
That means, the estimated effective subgroup under repeated simulations by BayRP is a set of subjects who belong to the effective subgroup in at least 90\% of the simulated trials. 

BayRP correctly identifies biomarkers 1 and 2 as important predictive biomarkers in all  scenarios. The second column of Figure \ref{fig:subgroup} shows the estimated  effective subgroup $\hat{\Delta}_{\mathrm{BayRP}}$ versus biomarkers 1 and 2, represented by the grid points in blue for  scenarios 1, 2, and 4. 
As shown in Figure  \ref{fig:subgroup}, the effective subgroups identified by the BayRP successfully recover the simulation truth. Scenario 3 is a NULL case, there is no effective subgroup. The BayRP identified $\hat{\Delta}_{\mathrm{BayRP}}=\emptyset$, which matches the simulation truth. 

For comparison, we implemented two alternative methods on subgroup identification for each simulated trial:  a Bayesian linear regression (LR) method and GUIDE \citep{loh2015regression}. 
Under the LR method,  the outcomes were modeled as a Bayesian linear regression considering all main effects and  interaction effects between   treatments and biomarkers: $$y_i\mid z_i, \xb_i = \beta_0 + \beta_1z_i + \balpha \xb_i + \bgamma z_i \xb_i + \epsilon_i,$$ where $\epsilon_i\iid N(0, \sigma^2)$. We assumed non-informative conjugate priors, $1/\sigma^2\sim \mathrm{Gamma}(0.1, 0.1)$ and $(\beta_0, \beta_1, \balpha,  \bgamma)\sim MN(0, 20{\bm I})$, where $\mathrm{Gamma}(a, b)$ denotes a gamma distribution with mean $a/b$, $MN(\bmu, \Sigma)$ denotes a multivariate normal distribution with mean $\bmu$ and covariance $\Sigma$, and ${\bm I}$ denotes the identity matrix. 
The posterior samples were obtained by a Gibbs sampling procedure. Since LR did not provide a formal way to report the effective subgroup under repeated simulations, we estimated the effective subgroup under LR following \eqref{eq:ASIED} under BayRP for a fair comparison. Specifically, we calculated the treatment effect $q_d^{h, (b)}$ for each grid $d$ with biomarker $\xb_d$ at MCMC iteration $b$ in trial $h$ using the predicted values of LR under two treatments, that is 
 $q_d^{h, (b)}=E^{h, (b)}(y_d\mid z_d=2, \xb_d) - E^{h, (b)}(y_d\mid z_d=1, \xb_d)$. We determined that the grid $d$ with biomarker $\tilde{\xb}_d$ belonged to the effective subgroup in trial $h$ if 
$\frac{1}{B}\sum_bI(q_d^{h, (b)}\geq \mathrm{LRV})>\xi.$ 
Similar to \eqref{eq:ASIED}, we then defined the estimated effective subgroup $\hat{\Delta}_{\mathrm{LR}}$ under LR as a collection of subjects who belong to the effective subgroup in at least 90\% of the simulated trials. 
Under the GUIDE method, we downloaded GUIDE software (\url{http://www.stat.wisc.edu/~loh/guide.html}, version 27.3) and applied the ``Gi" algorithm with the default parameter setting  to construct 0-SE least-square regression tree. All   biomarkers as well as the treatment assignments were used to construct the tree. We ran GUIDE to the simulated datasets and defined the effective subgroup as the set of subjects among all grid points whose predictive treatment effects are larger than LRV. Then $\hat{\Delta}_{\mathrm{GUIDE}}$ under GUIDE can be estimated in the same way as $\hat{\Delta}_{\mathrm{LR}}$. 

For the null scenario 3, both LR method and GUIDE successfully recover the simulated truth, $\hat{\Delta}_{\mathrm{LR}}=\hat{\Delta}_{\mathrm{GUIDE}}=\emptyset$.
The third and fourth columns of Figure  \ref{fig:subgroup} plot the estimated effective subgroup by the LR method and GUIDE in scenarios 1, 2, and 4, respectively. We can clearly see that LR cannot recover the simulated truths in these three scenarios when an effective subgroup exists. In contrast, the effective subgroups estimated by GUIDE have a large overlap with the simulated truths in scenarios 1 and 2, comparable to BayRP. However, in scenario 4 where there is one binary predictive biomarker (i.e., biomarker 1) and one continuous predictive biomarker (i.e., biomarker 2), GUIDE cannot identify any effective subgroup.

In addition, we report the true positive rate (TPR) and true negative rate (TNR) of the effective subgroup finding under BayRP, LR, and GUIDE. We define two quantities: 
1) TPR = $\sum_{\{d: \xb_d\in S^o\}} \sum_{h=1}^{100} I (\xb_d\in \hat{\Delta})/(|S^o|\times 100)$, where $|S^o|$ is the number of grid points in the simulated true effective subgroup; 2) TNR  = $\sum_{\{d: \xb_d\notin S^o\}} \sum_{h=1}^{100} I (\xb_d\notin \hat{\Delta})/(|\Omega \setminus S^o|\times 100)$. As shown in Table \ref{table:sub}, all three methods achieve high TNR, but the BayRP achieves much higher TPR compared to both the LR method and GUIDE, especially for scenario 4 where TPR is 0.87 for BayRP and 0 for either LR or GUIDE. Scenario 3 is not included since there is no effective subgroup. 

\begin{table}[h]
\centering
\begin{tabular}{c|ccc|ccc}
\hline
&&TPR&&&TNR\\
\hline
Scenario&BayRP &LR&GUIDE&BayRP&LR&GUIDE\\
\hline
1&0.97&0.54&0.95&1.00&1.00&0.99 \\
2&0.96&0.38&0.89&0.92&0.98&0.89 \\
4&0.87&0.00&0.00&1.00&1.00& 0.98 \\
\hline
\end{tabular}
\caption{The true positive rate (TPR) and true negative rate (TNR)  of the subgroup finding under BayRP, LR, and GUIDE.}
\label{table:sub}
\end{table}

In summary, the proposed BayRP can accurately identify the predictive biomarkers and the subgroup with enhanced treatment effect, building a solid foundation for the ASIED trial design. 
Note that the subgroup identification model is independent of the adaptive enrichment decision framework proposed in Section \ref{sec:design}, and any subgroup identification model such as random forest, SUBA, or BART could be implemented to the decision-making framework of the proposed ASIED.  This makes ASIED a flexible adaptive enrichment design that combines the tasks of ``subgroup identification" and ``enrichment." We choose BayRP since it is a robust Bayesian tree model but yields satisfactory performance when the sample size is small, unlike other complex models requiring large sample sizes for accurate estimation, such as SCUBA.

\section{ASIED Operating Characteristics}
\label{sec:oc}

In this section, we evaluate the performance of the proposed ASIED design on trial operating characteristics via simulation studies. In each trial, the maximum sample size was $N=180$  and subjects were equally randomized to the placebo or the investigational drug. We set LRV=2.37 and TV=3.08, following the rationale provided  in the motivating AD trial. The first interim analysis would be conducted after $n_1=100$ patients were enrolled. If a second interim analysis is needed, $n_2=40$. We assumed that $K=4$ biomarkers were available for each patient and $p(\nu_k)=1/5$, $k=0, \dots, 4$, indicating a uniform prior on the biomarker selection in the proposed BayRP on subgroup identification.

The decision-making framework depends on the parameters $\xi_1$ and $\xi_2$, which defines decisions ``Go", ``Stop", and ``Gray zone." 
A practical implementation of the proposed design should consider what would be acceptable decision risks to determine the 
tuning parameters $\xi_1$ and $\xi_2$. To this purpose, we define three acceptable risks at the first interim analysis: false stop risk (FSR), false go risk (FGR), and false enrich risk (FER). FSR is an acceptable risk that a ``Stop" decision is made at the first interim analysis when the truth is that there exists an effective subgroup or the all-comer is effective. FGR is an acceptable risk that a ``Go" or ``Gray zone" decision is made when the truth is that there is no effect in all-comers or in any subgroup. FER is an acceptable risk that an effective subgroup is enriched at the first interim analysis when the truth is that all-comers are effective. These risks could differ for different drugs and with the attitude to risk for different companies. In our case design, we set FSR=0.05, FGR=0.1, and FER=0.15. Note these values can be changed, as long as they can be justified.

To evaluate the performance of ASIED in operating characteristics, we assumed that all the biomarkers were continuous and generated from $\mathrm{Uniform}(-1, 1)$. The responses $y_i$'s were generated from $y_i=0.75 + \beta_0I(z_i=2) + \beta_1 I(x_{i1}>-0.4)I(z_i=2) + \epsilon_i$, where $\epsilon_i\sim N(0, 0.5^2)$. Here different values of $\beta_0$ and $\beta_1$ were selected to generate different scenarios.  %as shown in Table \ref{table:truth}. 
For example, when $\beta_0=0.25$ and $\beta_1=2.55$, there exists an effective subgroup $\Delta=\{i: x_{i1}>-0.4\}$ whose treatment effect is larger than LRV but smaller than TV. We considered five scenarios, as shown in Table \ref{table:truth}.

\begin{table}[h]
\centering
\begin{tabular}{cccc}
\hline
Scenario & $\beta_0$&$\beta_1$&Truth\\
\hline
1&$0.25$ & 2 &no effective subgroup\\
2&$0.25$ & 2.55  &effective subgroup($>$LRV, $<$TV)\\
3&$0.25$ & 2.83  &effective subgroup ($=$TV)\\
4&$2.6$ &  0 &effective all-comers ($>$LRV, $<$TV)\\
5& 3.08 & 0  &effective all-comers ($=$TV)\\
\hline
\end{tabular}
\caption{The simulation truths by different  values of $\beta_0$ and $\beta_1$.}
\label{table:truth}
\end{table}

\noindent{\bf Determining $\xi_1$ and $\xi_2$. } We first implemented ASIED design to the five   scenarios at the first interim analysis to determine $\xi_1$ and $\xi_2$ by controlling FSR, FGR, and FER. 
Table S1 in Supplement C presents the sensitivity analysis of the decisions for different combination values of $\xi_1$ and $\xi_2$. With the guidance of the three pre-determined risks, we set  $\xi_1=0.8$ and $\xi_2=0.1$. More details are discussed in Supplement C.

\begin{figure}[htbp]
\centering
\begin{tabular}{ccc}
\includegraphics[width=.33\textwidth,page=1]{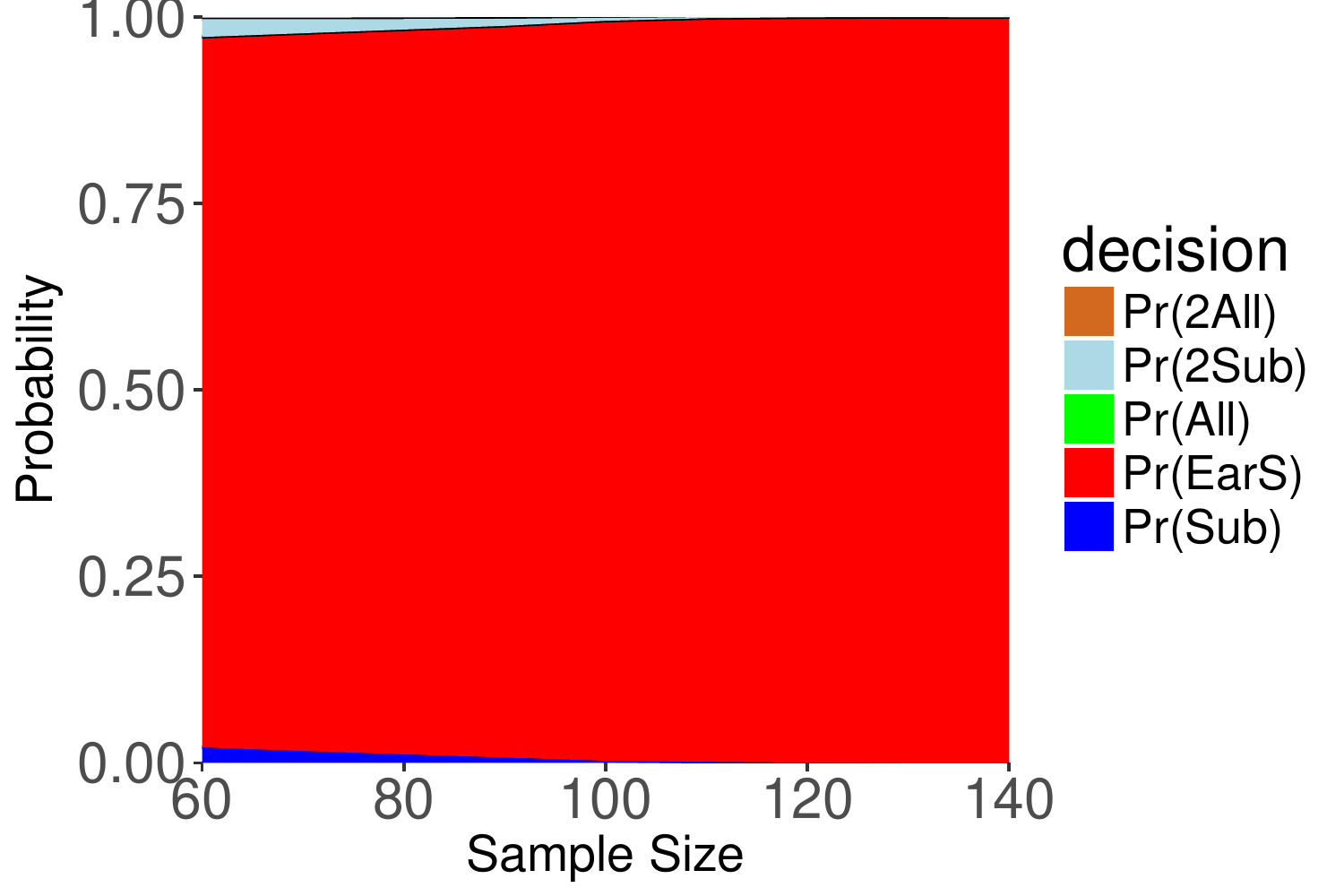} & \includegraphics[width=.33\textwidth,page=2]{./figs/sensitivity_ss} & \includegraphics[width=.33\textwidth,page=3]{./figs/sensitivity_ss} \\
(a) Scenario 1 & (b) Scenario 2& (b) Scenario 3\\
\end{tabular}
\begin{tabular}{cc}
\includegraphics[width=.33\textwidth,page=4]{./figs/sensitivity_ss} & \includegraphics[width=.33\textwidth,page=5]{./figs/sensitivity_ss}  \\
(c) Scenario 4 & (d) Scenario 5\\
\end{tabular}
\caption{Sensitivity analysis of the estimated probabilities of different decision outcomes at the first interim versus sample sizes. Pr(All) denotes the probability of continuing the trial with original population until the end of the trial. Pr(Sub) denotes the probability of continuing the trial with an enriched subpopulation until the end of the trial. Pr(EarS) denotes the probability of stopping the trial due to futility. Pr(2All) denotes the probability of continuing the trial with original population and plan a second interim analysis. Pr(2Sub) denotes the probability of continuing the trial with an enriched subpopulation and plan a second interim analysis. All probabilities are with respect to  100 repeated simulations.}
\label{fig:sen_ss}
\end{figure}

\noindent{\bf Sensitivity analysis to sample size at the first interim analysis. } To investigate the impact of sample size at the first interim on  the decision-making, we plot the estimated probabilities of different decisions  versus different sample sizes at the first interim analysis for all five scenarios in Figure \ref{fig:sen_ss}. %We considered four scenarios: scenario 1 (no effective subgroup), scenario 2 (effective subgroup with effect $>$LRV and $<$TV), scenario 4 (effective all-comers with effect $>$LRV and $<$TV), and scenario 5 (effective all-comers with TV effect). 
As shown in  Figure \ref{fig:sen_ss}, the sample size $n_1=100$ at the first interim analysis yields satisfactory decision outcomes. For instance, when there exists an effective subgroup (scenario 2), the probability of continuing the trial with an enriched subpopulation increases as the sample size increases. 

\noindent{\bf Operating characteristics. } Now we are ready to evaluate the performance of ASIED trial design in terms of operating characteristics. As described in Section \ref{sec:design}, ASIED has three possible final recommendations, with $a = 2$ denoting a recommendation of investigational drug for all-comers; $a = 1$ denoting a recommendation of investigational drug for a subgroup; and $a = 0$ denoting no recommendation for investigational drug. If the trial is stopped early due to futility, ASIED records the terminal decision $a = 0$. 
We implemented the ASIED  to the five scenarios shown in Table \ref{table:truth}. 
Table \ref{table:op} reports the probabilities of the five interim decisions and 
three final Go/Stop recommendations for each scenario under 100 repeated simulations. 
As shown in Table \ref{table:op}, ASIED achieves desirable operating characteristics. For example, when $\beta_0=0.25, \beta_1=2.55$, we recommend the investigational drug for an effective subgroup with probability 0.96 and recommend stop for the drug with probability 0.04; when $\beta_0=0.25$ and $\beta_1$ increases to 2.83, we recommend the investigational drug for an effective subgroup with probability 0.94 and for all-comers with probability 0.06. 

\begin{table}[h]
\centering
{\small
\begin{tabular}{ccccccc}
\hline
&&&&Interim Analysis &\\
Scenario&Truth&Pr(All)&Pr(Sub)&Pr(EarS)&Pr(2All)&Pr(2Sub)\\
\hline
1 & no effective subgroup & 0 &0.01&{\bf 0.99}&0&0\\
2 & effective subgroup ($>$LRV, $<$TV) &0&{\bf 0.96}&0.04&0&0\\
3 & effective subgroup ($=$TV)&0.1&{\bf 0.9}&0&0&0\\
4 & effective all-comers ($>$LRV, $<$TV)&{\bf 0.95}&0&0.05&0&0\\
5 & effective all-comers ($=$TV)&{\bf 1}&0&0&0&0\\
\hline
\end{tabular}
\begin{tabular}{ccccc}
\hline
&&&Final Recommendation of Go/Stop\\
Scenario&Truth&Pr($a$=2)&Pr($a$=1)&Pr($a$=0)\\
\hline
1 & no effective subgroup& 0&0&{\bf 1}\\
2 & effective subgroup ($>$LRV, $<$TV) &0&{\bf 0.96}&0.04\\
3 & effective subgroup ($=$TV)&0.06&{\bf 0.94}&0\\
4 & effective all-comers ($>$LRV, $<$TV)&{\bf 0.95}&0&0.05\\
5 & effective all-comers ($=$TV)&{\bf 1}&0&0\\
\hline
\end{tabular}  }
\caption{The operating characteristics of ASIED.  }
\label{table:op}
\end{table}

%
%Lastly, we study the ASIED efficacy by comparing the proposed enrichment design with the design without the enrichment procedure. That means, we do not modify the study entry criteria during the interim analysis. We call this design the WO design. Denote $y_i^h$ and $z_i^h$ to be the response and treatment assignment for patient $i$ in the $h^{th}$ simulated trial, $h=1, \dots, 100$, 
%we define the overall mean treatment effect (OMT) as $\mathrm{OMT} = \frac{1}{100} \sum_{h=1}^{100} \big(\frac{\sum_{i=n+1}^N y_i^hI(z_i^h=2)}{\sum_{i=n+1}^N I(z_i^h=2)} - \frac{\sum_{i=n+1}^N y_i^hI(z_i^h=1)}{\sum_{i=n+1}^N I(z_i^h=1)}\big)$, which is the mean response differences between the investigational drug and the placebo  after the interim analysis. As shown in Table \ref{table:oc}(b), the ASIED yields higher OMT compared to the WO design. 

\section{Conclusion}
\label{sec:con}

We develop a Bayesian random partition (BayRP) model  to search for   subgroups with enhanced treatment effect  and a novel enrichment design, ASIED, with a robust decision-making framework that either allows the study to continue with all-comers or adapt to a subpopulation with enhanced treatment effect based on the  interim accumulated data. 
BayRP is fully Bayesian, providing principled and coherent inference on the effective subgroup identification. As shown in the simulation  studies, BayRP is flexible and capable of recovering the  effective subgroups with various types of biomarkers. The rule for effective subgroup report is also flexible, as we can tune the confidence level $\xi$ in \eqref{eq:sub} depending on the goal of the study. 
%Through simulation studies on the operating characteristics of ASIED, we have demonstrated the importance of identifying subgroups and modifying the study entry criteria in an adaptive enrichment design when there exist subgroups with enhanced treatment effects. 

More importantly, to the best of our knowledge, ASIED represents the first   attempt in the literature to create a set of comprehensive rules to guide enrichment decisions at interim in the framework where  a multilevel target profile (LRV and TV) is incorporated.  The merit of using inferred treatment effect and probability statement to make enrichment decisions at interim or decisions about the drug's next step development at final is that not only the estimated treatment effect but also the associated variability are taken into consideration in the decision making. Simulation studies with sensitivity analyses show that ASIED achieves desirable operating characteristics. 
The decision-making criteria proposed in this paper can be tailored based on users' risk tolerance. 
For example, users can calibrate the hyperparameters $\xi_1$ and $\xi_2$ by their own preferences on risks such as false go risk, false  stop risk, etc. This is critical to precision medicine.

There are several potential extensions. First,  functionality of BayRP can be extended to include an algorithm that can reduce the dimension of the covariate space. This will make the subgroup identification more powerful and efficient when a large number of potential predictive biomarkers are available. 	Second, we can expand BayRP to cover various types of outcomes. For example, for survival outcomes we only need to change the sampling model $p(y\mid z_i, \Theta, \Pi)$.  
%Second,   the proposed ASIED focuses on the subgroup identification and study entry criteria enrichment \xx 
%at the interim analyses. \rr Question about the next statement: when the in the design that other treatments are involved? It is not clear. \xx One could easily add to the ASIED an adaptive patient allocation algorithm to assign patients to their superior treatments and 
%a final recommendation of a suitable patient population for a follow-up trial. 
%\rr (Second, when a large number of biomarkers are available, 
%we can extend the BayRP to incorporate variable selection. In addition, we consider various types of responses in BayRP, although the proposed model can be easily extend to other types of outcomes. For example, for survival outcomes, we only need to change the sampling model $p(y\mid z_i, \Theta, \Pi)$.\xx  
Lastly, we can extend the ASIED   to accommodate missing responses and delayed responses in the case of having an outcome that takes a relatively long time to be observed.

\section*{Acknowledgement}
This work is funded by the MedImmune L.L.C. and Johns Hopkins University collaboration project. We thank Prof. Constantine Frangakis for his useful discussion on the enrichment design.

\bibliographystyle{biom}
\bibliography{reference}

\end{document}